\documentclass[epj]{svjour}
\usepackage{graphicx}
\usepackage{amsmath}
\usepackage{amssymb}
\usepackage{dcolumn}
\usepackage{mathtools,cuted}
\newcommand{\AlTwoOThree}{Al$_{\mathrm{2}}$O$_{\mathrm{3}}$}
\newcommand{\MgO}{MgO}

\newcommand{\SiOTwo}{SiO$_{\mathrm{2}}$}
\newcommand{\phdag}{{\phantom{\dag}}}
\begin{document}
%
%\title{Electron kinetics across plasma-wall interfaces}
%\title{Probing the electron kinetics at the plasma interface}
\title{Electron kinetics at the plasma interface}
\author{%F. X. Bronold \and M. Pamperin \and E. Thiessen \and H. Fehske
F. X. Bronold \and H. Fehske \and M. Pamperin \and E. Thiessen 
}                     % Do not remove
\institute{Institut f\"ur Physik, Ernst-Moritz-Arndt-Universit\"at Greifswald, D-17489 Greifswald, Germany}
\date{Received: date / Revised version: date}
\abstract{
The most fundamental response of an ionized gas to a macroscopic object 
is the formation of the plasma sheath. It is an electron depleted space 
charge region, adjacent to the object, which screens the object's 
negative charge arising from the accumulation of electrons from the 
plasma. The plasma sheath is thus the positively charged part of an 
electric double layer whose negatively charged part is inside the wall. 
In the course of the Transregional Collaborative Research Center SFB/TRR24 
we investigated, from a microscopic point of view, the elementary 
charge transfer processes responsible for the electric double layer 
at a floating plasma-wall interface
and made first steps towards a description of the negative part of the 
layer inside the wall. Below we review our work in a colloquial manner, 
describe possible extensions, and identify key issues which need to be 
resolved to make further progress in the understanding of the electron 
kinetics across plasma-wall interfaces. 
 % end of PACS codes
} %end of abstract
\maketitle

\section{Introduction}
\label{Introduction}

%plasma sheath~\cite{Robertson13,Riemann91,SB90}
%emissive sheath~\cite{SKR09,CU16}
%solid-based integrated microdischarges~\cite{OE05,WTE10,DOL10}
%electric double layer between two gaseous plasmas~\cite{AA71,SB83,Charles07,Raadu89}

The most fundamental manifestation of the interaction of a solid surface with an ionized gas
is the formation of an electric double layer consisting, respectively, on the plasma and the
solid side of the interface of an electron-depleted and an electron-rich space charge region.
It arises because electrons are deposited more efficiently inside or on top of the solid,
depending on the electronic structure, than they are extracted from it by the neutralization
of ions and the de-excitation of heavy neutral particles. That an electric double layer forms
at a plasma-facing solid (plasma interface), having a negative part inside the solid and
a positive part inside the plasma, is known since the beginnings of gaseous electronics~\cite{LM24}.
A microscopic understanding of the charge transfer across the plasma interface on par with 
an understanding of charge transfer between two gaseous plasmas~\cite{AA71,SB83,Charles07,Raadu89}
or between two solids, such as, Schottky contacts~\cite{Tung01} is however still 
lacking. What has been studied in depth so far is only the merging of the plasma-based part of 
the double layer--the plasma sheath--with the quasi-neutral bulk 
plasma~\cite{Robertson13,Franklin03,Riemann91} and how it is affected by the emissive properties
of the surface, that is, by electron/ion reflection and electron emission~\cite{CU16,SKR09,HZ66}.

As far as the theoretical description of this fundamental electronic response of the plasma
interface is concerned it is usually assumed that the processes inside the solid
occur on spatial/temporal scales too small/fast to affect the plasma~\cite{Franklin76}. For 
the plasma species the solid is thus assumed to be only a reservoir characterized by surface 
parameters, such as, absorption, reflection, and emission probabilities, which have to be 
obtained either by quantum-mechanical calculations or by separate measurements. Encapsulating
the physics of charge transfer across the plasma interface into a set of surface parameters,
that is, considering electron absorption, reflection, and extraction as elementary surface 
collision processes has a long tradition in plasma physics (see~\cite{USB00,GMB02,KDK04,Kushner05,LS09,SEK13} 
for representative recent applications of this philosophy). Usually the surface parameters 
are assumed to be just numbers, independent of energy and angle, which in general is of course
not true. The crude description reflects simply the fact that in most cases the surface parameters
have been neither worked out theoretically nor measured experimentally. In particular at low
impact energies the data base is rather sparse or outdated as emphasized for the particular case 
of secondary electron emission by Tolias~\cite{Tolias14a}. It is only recently that experimental
work started anew to determine at energies relevant for low-temperature plasma applications 
surface parameters, such as, the electron backscattering probability~\cite{DAK15} or the secondary 
electron emission coefficient~\cite{DBS16,MCA15}. 

Characterizing charge transfer across the plasma interface by a set of (energy- and angle-dependent)
surface parameters is justified as long as the scales of charge transport/relaxation inside the plasma 
and inside the solid are well separated. In situations, however, where they approach each other, as we 
expect it to occur soon in arrays of integrated microdischarges~\cite{KSO12,DOL10}, because of the 
continuing miniaturization efforts in this field~\cite{EPC13}, it is no longer sufficient to treat 
electron deposition and extraction across the plasma interface as elementary surface collision 
processes. Instead it will be necessary to describe the charge transport across the interface 
selfconsistently with the charge dynamics on both sides of it. Such an approach is also required if 
one wants to understand in detail how the electronic non-equilibrium of the plasma is transferred 
to the solid, or if the physical system of interest consists of a plasma and a solid component, as 
it is the case for the plasma bipolar junction transistor~\cite{WTE10}.  

Motivated primarily by the prospects of solid-based opto-electronic plasma 
devices~\cite{KSO12,DOL10,EPC13,WTE10}, but also with an eye on dielectric barrier 
discharges~\cite{PS15,TBW14,BWS12}, we initiated in the framework of the Transregional Collaborative 
Research Center  SFB/TRR24 a still on-going effort to understand the charge transport, that is, the
electron kinetics across the 
plasma interface from a microscopic point of view. Part of our work is devoted to the calculation
of surface parameters (which may be also functions of energy and angle). Besides assisting future 
experimental efforts to measure them the parameters will help to make the modeling of the
plasma-wall interaction more realistic. Considering electron deposition and extraction
as elementary surface collision processes we calculated for dielectric surfaces probabilities
for electron sticking/backscattering~\cite{BF17a,BF15,HBF11,HBF10b,HBF10a} and secondary 
electron emission~\cite{MBF12b,MBF12a} and investigated how electronic correlations affect the 
neutralization of ions on metallic surfaces~\cite{PBF15b,PBF15a}.
In addition we started to explore the fate of the surplus electrons inside the wall, 
that is, the solid-based negative part of the electric double layer~\cite{BF17b,HBF12a} and 
proposed a method to measure the charge of a dielectric particle embedded in a plasma optically 
by Mie scattering~\cite{THB14,HBF13,HBF12b}. In the following sections we colloquially review 
our work, present numerical results for material systems not considered in our previous 
work, or of exploratory calculations extending it, and identify key issues which need to 
be resolved for making further progress in the microscopic understanding of charge 
transfer across plasma-wall interfaces.

\section{Electron absorption and backscattering}
\label{Absorption}

The interaction of electrons with surfaces is central for a great variety of surface 
diagnostics as well as materials processing techniques. It has been studied in great
detail (see references in~\cite{BF17a,BF15}). This knowledge however is not of immediate use 
for the modeling of electron-surface interaction in plasma applications, such as, dielectric 
barrier discharges, Hall thrusters, or the divertor region of fusion plasmas. The reason is the 
difference in the electron energies. Whereas in surface diagnostics, for instance, in 
electron microscopy, the energy of the electron probing the surface is at least 100~eV the energy 
of electrons hitting the wall of a bounded plasma is typically less than 10 eV. In this energy 
range little is known quantitatively about the interaction of electrons with surfaces. The
probability with which a low energy electron gets stuck in the surface and hence contributes to 
the solid-based negative part of the electric double layer is essentially unknown for the
materials used as plasma walls. Very often it is thus assumed the probability is unity
(perfect absorber assumption), irrespective of the angle of incidence and the wall material.
An electron, however, impinging on a solid surface is either reflected, inelastically 
scattered or temporarily deposited to the surface with possible trapping states (or sites) 
depending on the electron affinity of the surface. The perfect absorber model can thus not be 
universally valid.

To overcome the limitations of the perfect absorber model we initially modelled electron 
trapping/sticking at low energies as a physisorption process in the surface's image potential using 
rate equations for the occupancies of the image states~\cite{HBF11,HBF10b,HBF10a} and relying on 
methods developed for describing the adsorption of neutral atoms~\cite{KG86,GKT80}. This approach,
applicable to electro-negative dielectrics, such as LiF, yields very small sticking probabilities. 
Due to lack of experimental data it is presently unclear how realistic these results are.
We also investigated electron absorption by electro-positive dielectrics~\cite{BF17a,BF15}, 
such as, \AlTwoOThree\, or \SiOTwo. Since they are more commonly used in plasma applications 
we discuss the approach applied to them in more detail.

An electron hitting a dielectric with positive electron affinity may enter, after a successful 
transmission through the surface potential, the conduction band of the dielectric.
Inelastic scattering due to phonons may then push the electron back to the interface and, 
after a successful transmission through the surface potential in the reverse direction, back 
to the plasma. In general, the probability for this chain of events is finite. Hence, the sticking 
probability should be less than unity with the particular numerical value depending on the
electronic structure of the surface and the efficiency of the scattering processes. 

\begin{figure*}[t]
\begin{minipage}{0.38\linewidth}
\includegraphics[width=\linewidth]{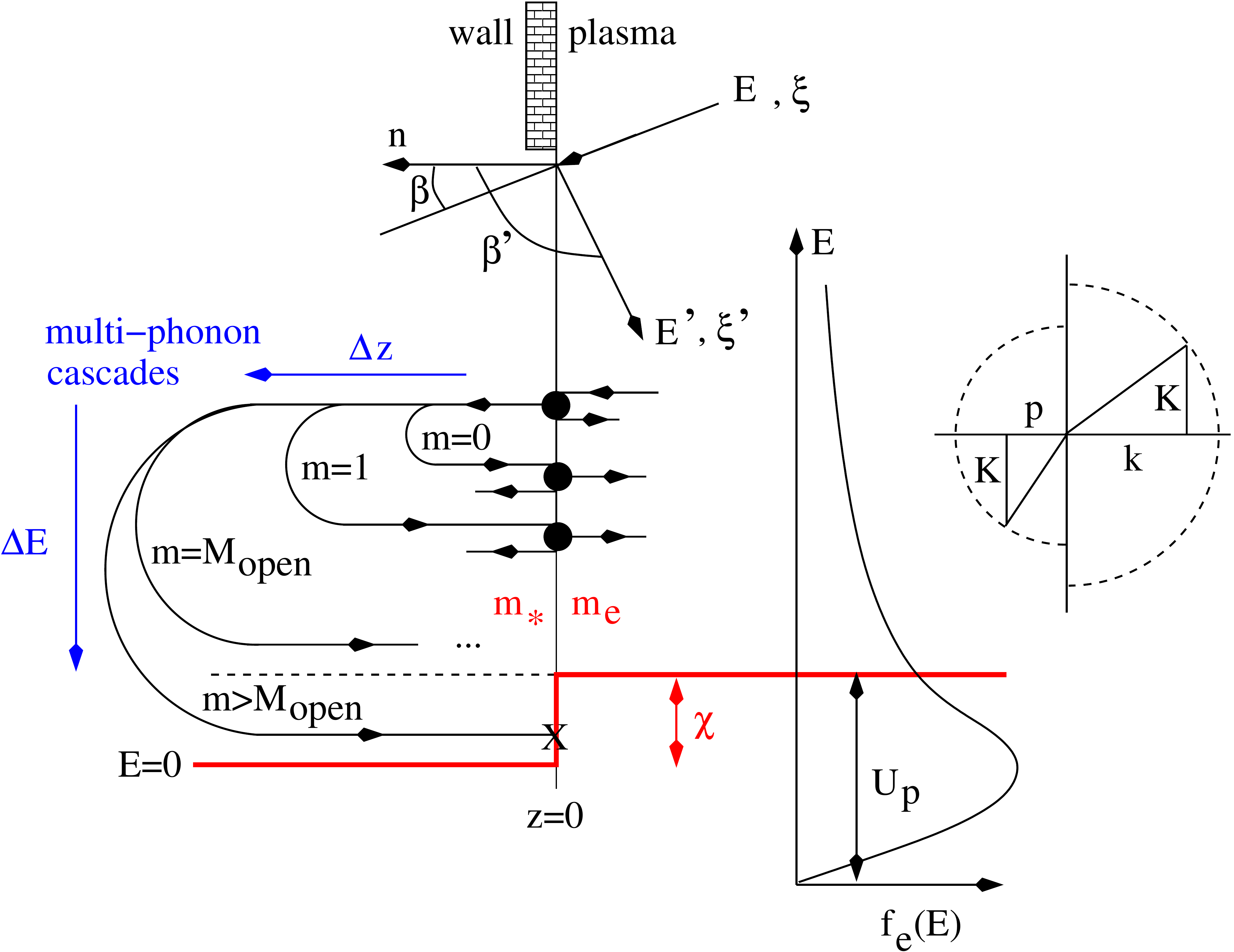}
\end{minipage}\begin{minipage}{0.31\linewidth}
%\rotatebox{270}{\includegraphics[width=0.84\linewidth]{figs/SmAl2O3.pdf}}
\rotatebox{270}{\includegraphics[width=0.84\linewidth]{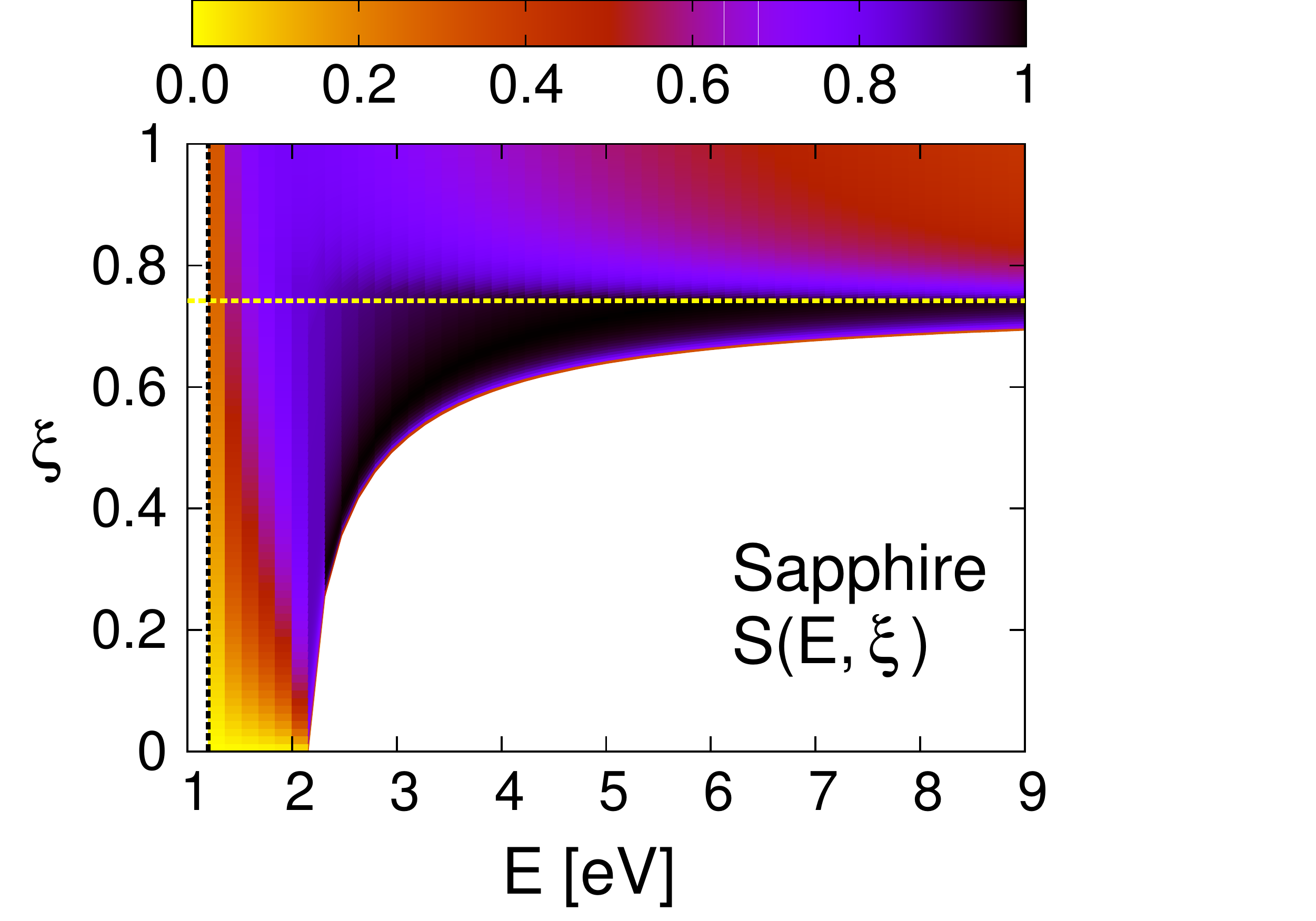}}
\end{minipage}\begin{minipage}{0.31\linewidth}
%\rotatebox{270}{\includegraphics[width=0.84\linewidth]{figs/P1mAl2O3.pdf}}
\rotatebox{270}{\includegraphics[width=0.84\linewidth]{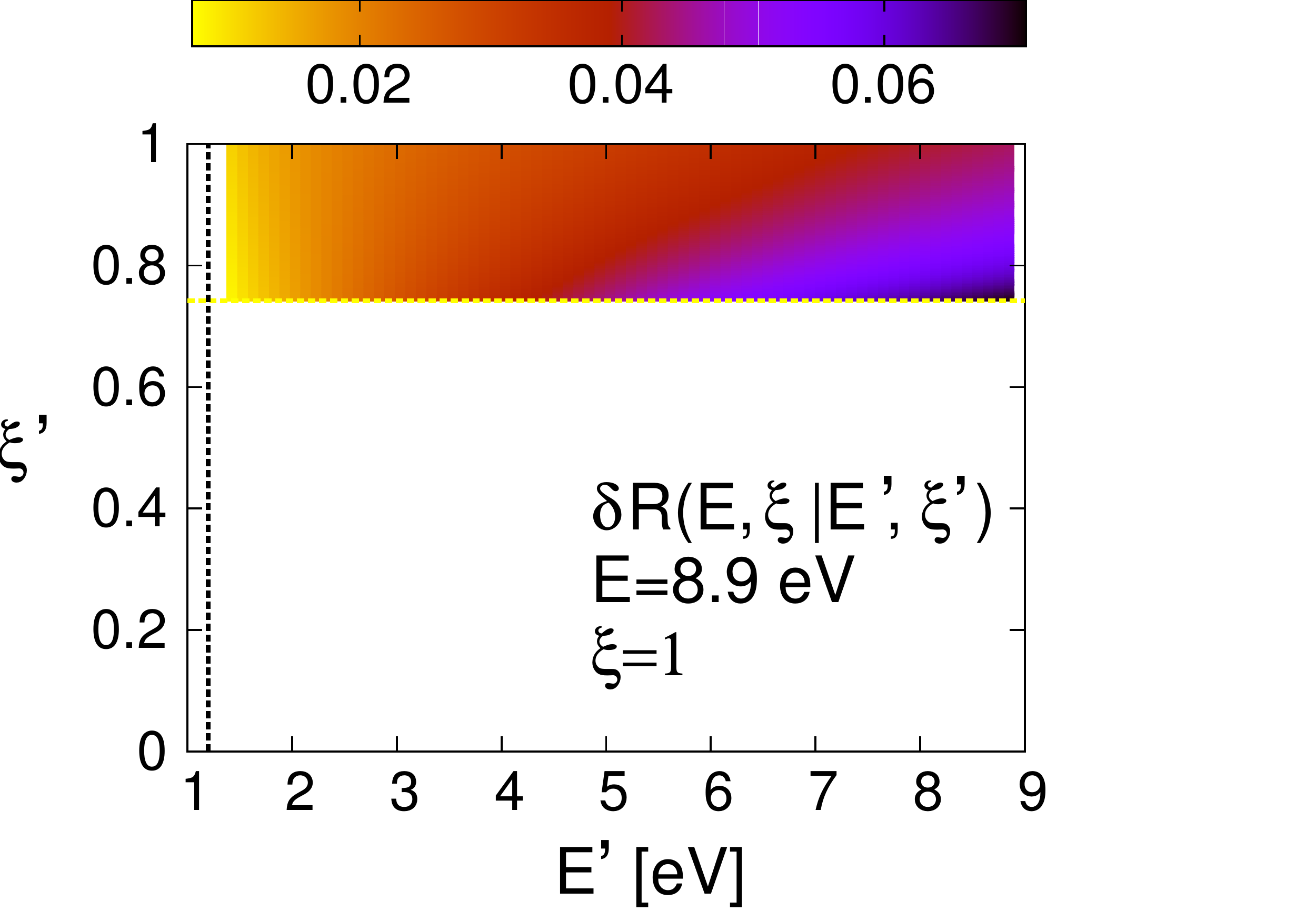}}
\end{minipage}
\caption{(Color online)
Left panel: Interface model and principal idea of the calculational approach
used for studying electron absorption/backscattering by dielectric walls
with positive electron affinity $\chi>0$. The direction cosines inside ($\eta$) 
and outside ($\xi$) the surface can be deduced from the two semicircles. 
Central panel:  Angle-resolved
electron sticking probability $S(E,\xi)$ for \AlTwoOThree\, (sapphire) with
material parameters $m_e/m_*$, $\chi$, $E_g$, and $\varepsilon$ as listed in 
Table~\ref{MaterialParameters}. Below the
yellow dashed line, indicating  $\xi=\sqrt{1-\overline{m}_e}$, inelastic
backscattering has no effect on the sticking probability. In the white region
$S(E,\xi)=0$ due to total reflection of the electron by the potential step
mimicking the surface potential. Right panel: The function
$\delta R(E\xi|E^\prime\xi^\prime)$, describing diffuse backscattering of an
electron with energy $E$ and direction cosine $\xi$ into a state with energy
$E^\prime$ and direction cosine $\xi^\prime$, for an electron hitting the
\AlTwoOThree\ surface with $E=8.9\,{\rm eV}$ and $\xi=1$. Below the dotted
yellow line $\delta R(E\xi|E^\prime\xi^\prime)=0$ because diffuse backscattering
cannot lead to post-collision cosines $\xi^\prime < \sqrt{1-\overline{m}_e}$.
}
\label{StickBack}
\end{figure*}

Utilizing the electron's large penetration depth at the energies of a few electron volts~\cite{Cazaux12},
typical for plasma applications, we showed that the chain of events described in the previous paragraph 
gives rise to a sticking probability $S(E,\xi)$ which is the product of the probability ${\cal T}(E,\xi)$ for
quantum-mechanical transmission through the surface potential and the probability to stay inside the surface 
despite of inelastic backscattering inside it~\cite{BF17a,BF15}. To make the connection between absorption 
by and backscattering from the dielectric surface explicit, we recast the expression for $S(E,\xi)$ in the 
form 
\begin{align}
S(E,\xi) &= 1-\int_\chi^E dE^\prime \int_0^1 d\xi^\prime R(E\xi|E^\prime\xi^\prime) ~,
\end{align}
where $R(E\xi|E^\prime\xi^\prime)$ encodes the backscattering of an electron hitting the surface with 
energy $E$ and direction cosine $\xi$ to a state with energy $E^\prime$ and direction cosine $\xi^\prime$
(see Figure~\ref{StickBack} for the definition of the cosines and energies). It consists of two parts, 
\begin{align}
R(E\xi|E^\prime\xi^\prime) &= R(E,\xi) \delta(E-E^\prime)\delta(\xi-\xi^\prime) \nonumber\\
                           &+ \delta R(E\xi|E^\prime\xi^\prime)~,
\end{align}
with $R(E,\xi)=1-{\cal T}(E,\xi)$ the probability for quantum-mechanical reflection by the 
surface potential and 
\begin{align}
\delta R(E\xi|E^\prime\xi^\prime) &= \frac{\partial\eta^\prime}{\partial\xi^\prime} {\cal T}(E,\xi)
                    \rho(E^\prime){\cal B}(E\eta|E^\prime\eta^\prime) {\cal T}(E^\prime\xi^\prime) \nonumber\\
                                  &\times \theta(\xi^\prime-\sqrt{1-\bar{m}_e})
\end{align}
with
\begin{align}
{\cal B}(E\eta|E^\prime\eta^\prime)=
\frac{Q(E\eta|E^\prime\eta^\prime)}
{\int_0^1 d\eta^\prime \int_0^E dE^\prime \rho(E^\prime)Q(E\eta|E^\prime\eta^\prime)}
\label{Bfct}
\end{align}
the probability for diffuse backscattering. The differential coefficient 
$\partial\eta^\prime/\partial\xi^\prime$ 
arises because the direction cosines inside ($\eta$ and $\eta^\prime$) and outside ($\xi$ and $\xi^\prime$) 
the surface are different because of the mismatch between the electron masses and the three-dimensional 
potential step mimicking the surface potential. Due to conservation of the total energy $E$ and the lateral 
momentum $\vec{K}$ they are connected by 
\begin{align}
1-\eta^2 = \frac{E-\chi}{\overline{m}_e E}\big(1-\xi^2\big)~
\label{etaxi}
\end{align}
with $\bar{m}_e=m_*/m_e$ the mass of a conduction band electron in units of the bare electron mass and
$\chi>0$ the electron affinity of the dielectric. Using this relation for the post-collision 
direction cosines yields the relation $\partial\eta^\prime/\partial\xi^\prime$. 

For the model shown in Figure~\ref{StickBack} the transmission probability is given by 
\begin{align}
{\cal T}(E,\xi)=\frac{4\overline{m}_e k p}{(\overline{m}_e k + p)^2}
\label{Trans}
\end{align}
with $k=\sqrt{E-\chi}\,\xi$ and $p=\sqrt{\overline{m}_e E}\,\eta$ the $z-$components
of the electron momenta outside and inside the wall. In all formulae atomic units are used, 
that is, energy is measured in Rydbergs and lengths in Bohr radii. Since the signs of $k$ 
and $p$ are always the same in~\eqref{Trans} we define the direction cosines
$\xi$ and $\eta$ referenced, respectively, to the electron momenta outside and
inside the wall, by their absolute values: $\xi=|\cos\beta|$ and $\eta=|\cos\theta|$. 
This choice simplifies also the calculation of the function $Q(E\eta|E^\prime\eta^\prime)$ 
which is the essential part of $\delta R(E\xi|E^\prime\xi^\prime)$, the function 
describing diffuse backscattering by the surface. 

The function $Q(E\eta|E^\prime\eta^\prime)$ is proportional to the probability for an electron
penetrating the surface in a state with (total) energy $E$ and direction cosine $\eta$ to
backscatter after an arbitrary number of internal scattering events towards the interface in a
state with energy $E^\prime$ and direction cosine $\eta^\prime$. It is determined by a recursion
relation derived from the principle of invariant embedding~\cite{Dashen64,Vicanek99,GP07}.
The invariant embedding principle is particularly well suited for our purpose because it 
enables us to perform an expansion in the number of backscattering events which converges 
very fast because backscattering is usually much less likely than forward scattering. In fact, 
we can truncate the expansion already after one backscattering event. What is an advantage 
for the invariant embedding approach is a serious drawback for Monte Carlo simulations. 
Sampling rarely occurring backscattering trajectories is computationally demanding. Standard
Monte Carlo techniques are thus not particularly well-suited for the investigation of 
electron backscattering.

In~\cite{BF17a,BF15} we derived the recursion relation for the function $Q(E\eta|E^\prime\eta^\prime)$ 
under the assumption that the dielectric surface is at room temperature and the impinging 
electron has initially an energy $E<E_g$, where $E_g$ is the band gap of the dielectric. The first 
assumption implies that the scattering occurs primarily due to emission of optical phonons 
while the second assumption guarantees that Coulomb-driven electron energy dissipation can be 
ignored. Taking then into account that a forward scattering event leads in a very good 
approximation only to an energy loss $\omega$, where $\omega$ is the energy of the phonon, 
leaving the propagation direction of the electron however unchanged, 
$Q(E\eta|E^\prime\eta^\prime)$ can be written as 
\begin{align}
Q(E\eta|E^\prime\eta^\prime)=\sum_{m=0}^{M_{\rm open}} Q^1_{m}(E;\eta|\eta^\prime)
\delta(E-E^\prime-\omega^1_m)
\label{Qexpansion}
\end{align}
with $M_{\rm open}$ the number of forward scattering events at most possible for the 
initial energy $E$, $\xi$ the initial direction cosine, and $\omega^1_m=(1+m)\omega$. 
The expansion coefficients $Q^1_{m}(E;\eta|\eta^\prime)$ satisfy a linear recursion relation, 
to be found in~\cite{BF17a,BF15}, which can be solved numerically quite efficiently for given 
values of $E, \eta,$ and $\eta^\prime$. 

The physical picture behind our approach of calculating the electron sticking and 
backscattering probabilities for a dielectric surface with positive electron affinity
is illustrated in Figure~\ref{StickBack} together with representative results for an \AlTwoOThree\, 
surface. First let us consider the illustration on the left. It shows that for a dielectric
with positive electron affinity diffuse backscattering from the surface is due to an internal 
multi-phonon cascade consisting of a finite number of electron trajectories characterized 
by a single backscattering and an increasing number of forward scattering events. Each 
trajectory represents a backscattering channel contributing to the diffuse backscattering from
the surface if the electron moving along such a trajectory traverses the surface potential 
successfully from the solid to the plasma side. 

Let us now turn to the numerical data shown in the middle and the right panel of 
Figure~\ref{StickBack}. The middle panel gives $S(E,\xi)$ over the whole range of direction 
cosines $\xi$ and energies up to $9\,{\rm eV}$ above which Coulomb-driven scattering 
processes should be taken into account. In principle this is possible. Preliminary work 
indicated that the recursion relations, containing in this case also energy 
integrals, can still be solved efficiently by numerical means. Work in this direction is 
in progress but will not be reported here. The sticking probability $S(E,\xi)$ in 
the white region of the $(E,\xi)$-plane is identically zero. It is the region of total 
reflection at the three-dimensional potential step characterizing the surface potential. 
Below the yellow line, given by $\xi=\sqrt{1-\bar{m}_e}$, the sticking probability is given 
by the quantum-mechanical transmission probability, that is, $S(E,\xi)={\cal T}(E,\xi)$. 
This is a consequence of the mass mismatch $\bar{m}_e=m_*/m_e<1$ and the conservation of 
total energy and lateral momentum which force the perpendicular energy of an electron 
impinging on the surface with a direction cosine $\xi < \sqrt{1-\bar{m}_e}$ to  
be less than $\chi$, the depth of the surface potential, once it crossed the interface from 
the plasma side. Since at room temperature the electron cannot gain energy by absorbing 
phonons it is thus immediately confined to the surface once it entered it. Only above the 
yellow line, that is, for $\xi > \sqrt{1-\bar{m}_e}$ phonon emission may bring the electron 
back to the interface. The possibility of crossing it also from the solid side leads then 
in this part of the $(E,\xi)$-plane to $S(E,\xi)\le{\cal T}(E,\xi)$.

Having a numerical value for the sticking probability $S(E,\xi)$ is enough for determining 
for instance the charge of a dust particle in a plasma. If one is however interested in the 
feedback the surface has to the plasma the electron backscattering probability 
$R(E\xi|E^\prime\xi^\prime)$ is the more important quantity. 
It consists of the quantum-mechanical (specular) backscattering probability 
$R(E,\xi)$ and the diffuse backscattering probability $\delta R(E\xi|E^\prime\xi^\prime)$. 
The latter is shown in the right panel of Figure~\ref{StickBack} for an electron hitting an 
\AlTwoOThree\, surface with $E=8.9\,{\rm eV}$ and $\xi=1$. Similar plots arise for other 
values of $E$ and $\xi$. Below the yellow line, given now by $\xi^\prime=\sqrt{1-\bar{m}_e}$,
diffuse backscattering is not possible because the mass mismatch in conjunction with the 
conservation of total energy and lateral momentum prevent post-collision cosines 
$\xi^\prime<\sqrt{1-\bar{m}_e}$. Diffuse backscattering in these directions is thus 
impossible.  

The data presented in Figure~\ref{StickBack} are obtained for a laterally homogeneous surface.
Only then the lateral momentum is conserved leading to total reflection. In reality, surfaces 
in contact with a plasma are for sure laterally inhomogeneous. Lateral momentum is thus 
not conserved and total reflection suppressed. In~\cite{BF17a,BF15} we also considered this 
case, based on a simple model for interfacial scattering~\cite{SLN98} which destroys the lateral 
homogeneity. We then found astonishingly good agreement between calculated values for $S(E,\xi=1)$ 
and data obtained from electron-beam scattering experiments on MgO~\cite{CF62} and 
\SiOTwo~\cite{Dionne75} surfaces indicating that our approach captures the essential processes
responsible for low-energy electron backscattering from electro-positive dielectrics. 

So far we focused on electron energy relaxation due to scattering by optical phonons. But other
scattering processes can be included as well. The numerical effort increases, in particular, 
if Coulomb-driven scattering processes are included, such as, electron-hole pair generation due
to the impacting electron, becoming important for $E>E_g$. Still, we expect the approach 
to remain more efficient than Monte Carlo simulations. It should thus be a rather useful 
tool for analyzing the recent experimental data obtained for electron backscattering 
from plasma walls~\cite{DAK15} and for generating for various materials data for $R(E\xi|E^\prime\xi^\prime)$ 
and $S(E,\xi)$ to be used in plasma modeling.

\begin{table}[t]
  \begin{tabular}{|l|lllllll|}
    \hline
                  & $\frac{m_e}{m_h}$  & $\frac{m_e}{m_*}$  & $k_BT_*$      & $k_BT_h$ & $\chi$         & $E_g$      & $\varepsilon$ \\
                  &                    &                    & (eV)          & (eV)     & (eV)           & (eV)       &               \\\hline
    \SiOTwo       & 1.0                & 1.3                & 0.2           & 0.2      & 1.2            & 9.0        & 4             \\
    \AlTwoOThree  & 1.0                & 2.2                & 0.2           & 0.2      & 1.2            & 8.7        & 9.9           \\
    \MgO          & 1.0                & 2.5                & 0.2           & 0.2      & 1.2            & 7.8        & 9.8           \\\hline\hline
%                  &                    &                    &               &          &                &            &               \\
                  & $\frac{m_e}{m_i}$  &                    & $k_BT_e$      & $k_BT_i$ & $I$            & $\Gamma$   &               \\
                  & ($10^{-4}$)        &                    & (eV)          & (eV)     & (eV)           & (eV)       &               \\\hline
    ${\rm H}^+e$  & 5.4                &                    & 2.0           &  0.2     & 13.6           & 2.0        &               \\\hline
  \end{tabular}
\caption{Material parameters used in the calculations for the perfectly absorbing
collisionless plasma-wall interface. The image shift of the ion's ionization level
$I$ is neglected to keep the model as simple as possible. It should be also noticed
that the parameters give only an orientation. For real surfaces used in actual
experiments the values may deviate from the ones given due to materials science
aspects not addressed in this work.
}
\label{MaterialParameters}
\end{table}

\section{Electron extraction}
\label{Extraction}

Charge extraction by the neutralization of ions and/or the de-excitation of neutral particles
is an important elementary surface collision process because it may provide secondary electrons 
which in turn affect the overall charge balance of the plasma. At high impact energies electrons 
are ejected from the surface due to the transfer of the kinetic energy associated with the 
center-of-mass motion of the projectile to the surface (kinetic electron ejection) whereas 
at low impact energies electrons are released due to the transfer of the energy stored in 
the internal motion of the projectile's constituents (potential electron ejection)~\cite{PP99}.
In the SFB/TRR24 we were particularly interested in a theoretical description of potential 
electron ejection from plasma walls. Besides the well-known two-electron Auger processes 
(Ref.~\cite{MBF11} and below) we investigated secondary electron emission due to auto-detaching 
negative ions~\cite{MBF12b,MBF12a} and analyzed resonant neutralization of ions with an eye on 
mixed-valence and Kondo resonances~\cite{PBF15b,PBF15a}.

Our work on electron extraction from surfaces due to impacting atomic particles is based on 
multi-channel Anderson-Newns 
models~\cite{YM86} parameterized by experimental energies and target-projectile interactions 
deduced from considerations based on image charges~\cite{Gadzuk67a,Gadzuk67b}. The models 
act on the internal states of the projectiles taking their classical center-of-mass motion in 
the surface potential into account by time-dependent matrix elements. Recasting the models 
in terms of pseudo-particles~\cite{Coleman84}, representing the projectile's electronic 
configurations, as proposed in a pioneering work by Langreth and Nordlander~\cite{LN91},
quantum kinetic equations for their occurrence probabilities can be derived and numerically 
solved~\cite{LN91,SLN94a,SLN94b}. 

For the de-excitation of metastable ${\rm N}_2^*$ molecules on dielectric surfaces we 
obtained good agreement with experimental data~\cite{MBF12b,MBF12a}. The secondary electron 
emission coefficients we find for the neutralization of ${\rm He}^+$ 
on metal surfaces are also of the correct order of magnitude indicating that the 
semiempirical approach is working. From our point of view, the semiempirical approach 
is also the most suitable for describing elementary surface collision processes at plasma
walls which are in most cases not well enough characterized to justify a more elaborate 
theoretical description.

\begin{figure}[t]
%  \centering\includegraphics[width=0.9\linewidth]{figs/HeNeutralWithNegProcesses.pdf}
%  \includegraphics[width=\linewidth]{figs/AugerNeutralCartoon.pdf}
  \includegraphics[width=\linewidth]{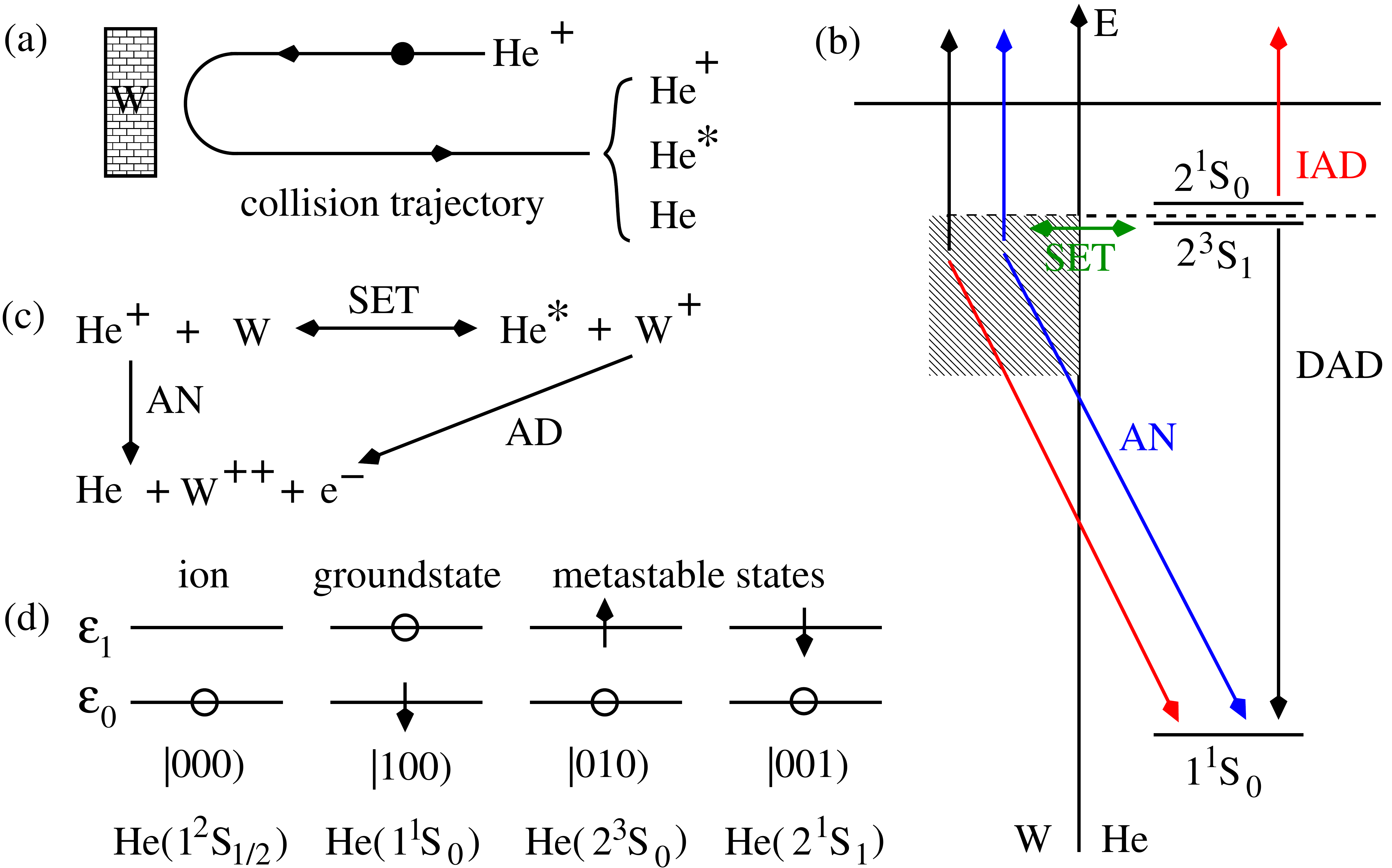}
  \caption{(Color online) Schematic representation of an 
   Anderson-Newns model based analysis of the neutralization of a helium ion on a tungsten 
   surface characterized by a Fermi energy $E_{\rm F}=6.4\,{\rm eV}$ and a work function
   $\Phi=4.5\,{\rm eV}$. Other
   projectile-target combinations and other collision scenarios can be treated in the same
   spirit. Panel (a) shows a collision trajectory leading to time-dependent matrix elements
   and thus to the necessity to use quantum-kinetic equations for the theoretical analysis of
   the neutralization process. The reaction channels included into our modeling are indicated
   in the (on scale) energy (b) and the channel (c) diagram. In the energy diagram the image  
   shifts are neglected for simplicity. The energy levels given for the projectile are the ones far 
   away from the surface. The helium ion may capture an electron from the metal by a single-electron  
   transfer (SET), changing its configuration from ${\rm He}^+{\rm (1s; 1^2{\rm S}_{1/2})}$ to
   either ${\rm He}^*{\rm (1s2s; 2^3{\rm S}_{1})}$ or ${\rm He}^*{\rm (1s2s; 2 ^1{\rm S}_0)}$,
   and subsequently Auger de-excite (AD) to ${\rm He}{\rm (1s^2; 1^1{\rm S}_0)}$ releasing
   thereby a secondary electron. The ${\rm He}^+{\rm (1s; 1^2{\rm S}_{1/2})}$ ion may however 
   also Auger neutralize (AN) to ${\rm He}{\rm (1s^2; 1^1{\rm S}_0)}$ which also leads to 
   the ejection of an electron. Panel (d) depicts the two-level system to which the electronic 
   configurations of the projectile involved in the collision process give rise to with 
   $\varepsilon_0$ the first ionization energy of helium ($24.6\, {\rm eV}$) and $\varepsilon_1$, 
   depending on the spin orientation, the ionization energy of the singlet ($4\, {\rm eV}$) or 
   triplet ($4.8\, {\rm eV}$) metastable state.}
  \label{HeNeutralProcesses}
\end{figure}

Although our prime motivation stems from the desire to characterize elementary surface 
collision processes at plasma walls we investigated also the neutralization of strontium ions 
on gold surfaces~\cite{PBF15b,PBF15a}, because experiments~\cite{HY10,HY11} indicated that 
it is perhaps driven by a mixed-valence resonance and thus by a feature which is paradigmatic 
for correlated quantum impurity systems. Our theoretical analysis could not fully support 
this conjecture but we found evidence for it deserving further exploration. The 
${\rm Sr}^+$:Au system is not relevant of course for plasma applications. It shows however 
that the charge transfer processes taking place at plasma interfaces can be embedded in a 
context making them interesting to a larger group of physicists.

To give a better idea of how electron ejection at low impact energies can be approached
by Anderson-Newns models we discuss secondary electron emission due to 
the neutralization of a ${\rm He}^+$ ion on a tungsten surface in more detail. For the 
purpose of this paper, which is to review our work performed in the SFB/TRR24, the 
discussion will stay at the qualitative level, mathematical details will be given 
elsewhere~\cite{PBF17b}. 

The essence of our approach is illustrated in Figure~\ref{HeNeutralProcesses}. Panel (a) 
of this figure shows a collision trajectory. For simplicity we assume the ${\rm He}^+$ 
ion to hit the surface perpendicularly. Hence, the position of the projectile in 
front of the surface is given by $z(t)=z_{\rm tp}+v|t|$, where $z_{\rm tp}$ is the
turning point, $v$ is the velocity of the projectile and $t$ runs from 
$-t_{\rm max}$ to $t_{\rm max}$ with $t_{\rm max}$ large enough to decouple the 
projectile from the target. Other collision angles can be also treated. At the level
of our models it just leads to a velocity-induced smearing out of the target's Fermi
edge~\cite{SHL03}. Below we present in fact data for grazing incident. Two main neutralization 
routes, put together in panels (b) and (c), are operative, each one leading to the ejection of 
an electron. The ion can either Auger neutralize (AN) to the groundstate of the atom or 
Auger de-excite (AD) to it in case a metastable configuration is formed before via a 
single-electron transfer. In that case a direct (DAD) and an indirect (IAD) channel exist. 
Assuming without loss of generality the electron of the ${\rm He}^+$ ion to have spin up, 
the electronic configurations involved in the neutralization processes summarized 
in panel (c) are ${\rm He}^+(1^2{\rm S}_{1/2})$ for the ion,  
${\rm He}(1^1{\rm S}_0)$ for the groundstate, 
${\rm He}^*(2^3{\rm S}_1)$ for the triplet metastable state, and 
${\rm He}^*(2^1{\rm S}_0)$ for the singlet metastable state. They give rise to 
the two-level system shown in panel (d). 

Instead of using Fermi operators acting on the two-level system we prefer to 
represent the electronic configurations of the projectile by pseudo-operators.
Recalling that the 1s and 2s shell of helium are involved in building up the 
electronic states of the projectile, we introduce operators $e^\dagger$, 
$s_{1\downarrow}^\dagger$, $s_{2\uparrow}^\dagger$, and $s_{2\downarrow}^\dagger$ 
by defining 
\begin{align}
\vert 000\rangle &= e^\dagger \vert {\rm vac}\rangle ~,~~
\vert 100\rangle = s_{1\downarrow}^\dagger \vert {\rm vac}\rangle~, \\
\vert 010\rangle &= s_{2\uparrow}^\dagger \vert {\rm vac}\rangle ~,~~
\vert 001\rangle = s_{2\downarrow}^\dagger \vert {\rm vac}\rangle~.
\end{align}
They denote, respectively, the ion ${\rm He}^+(1 ^2{\rm S}_{1/2})$, the 
groundstate atom ${\rm He}(1 ^1{\rm S}_0)$, and the triplet and singlet metastable states, 
${\rm He}^*(2 ^3{\rm S}_1)$ and ${\rm He}^*(2 ^1{\rm S}_0)$. Employing the 
reasoning of our previous work~\cite{MBF12b,PBF15b}, the Hamiltonian describing 
the neutralization of a ${\rm He}^+(1 ^2{\rm S}_{1/2})$ ion on a tungsten surface 
within the scenario summarized in Figure~\ref{HeNeutralProcesses} becomes 
\begin{align}
  H(t) &= \varepsilon^*_{s}(t) s_{2\downarrow}^\dagger s^\phdag_{2\downarrow}
  + \varepsilon^*_{t}(t) s_{2\uparrow}^\dagger s^\phdag_{2\uparrow}
  + \varepsilon_{g}(t) s_{1\downarrow}^\dagger s^\phdag_{1\downarrow} \nonumber\\
  &+ \sum_{\vec{q} \sigma} \varepsilon_{\vec{q} \sigma}(t) c^\dagger_{\vec{q} \sigma} c^\phdag_{\vec{q} \sigma}
  + \sum_{\vec{k} \sigma} \varepsilon_{\vec{k} \sigma} c^\dagger_{\vec{k} \sigma} c^\phdag_{\vec{k} \sigma} \nonumber\\
 &+ \sum_{\vec{k}\sigma} \big[ V^{\rm SET}_{\vec{k}\sigma}(t)c^\dagger_{\vec{k}\sigma} e^\dagger s^\phdag_{2\sigma} 
    + \text{H.c.} \big]\nonumber\\
 &+ \sum_{{\vec{k}_1}{\vec{k}_2}{\vec{k}^\prime}\sigma} 
    \big[V^{\rm AN}_{{\vec{k}_1}{\vec{k}_2}{\vec{k}^\prime\sigma\downarrow}}(t) c^\dagger_{{\vec{k}^\prime}\sigma}
   s^\dagger_{1\downarrow} \,e\, c^\phdag_{{\vec{k}_1}\downarrow} c^\phdag_{{\vec{k}_2}\sigma} + \text{H.c.} \big] \nonumber\\
 &+ \sum_{\vec{k}\vec{k}^\prime\sigma} \big[ V^{\rm DAD}_{\vec{k}\vec{k}^\prime\sigma}(t) 
   c^\dagger_{\vec{k}^\prime\sigma} c^\phdag_{\vec{k}\sigma}
   s^\dagger_{1\downarrow} s^\phdag_{2\downarrow} + \text{H.c.} \big]\nonumber\\
 &+ \sum_{\vec{k}\vec{q}\sigma} \big[ V^{\rm IAD}_{\vec{k}\vec{q}\sigma}(t) 
    c^\dagger_{\vec{q}\sigma} c^\phdag_{\vec{k}\downarrow} 
    s^\dagger_{1\downarrow} s^\phdag_{2\sigma} + \text{H.c.} \big] ~,
\label{ANM}
\end{align}
where the operator $c_{\vec{k}\sigma}^\dagger$ creates an electron with momentum $\vec{k}$ and
spin $\sigma$ in the conduction band of the metal and the operator $c_{\vec{q}\sigma}^\dagger$ 
puts an electron in an unbound projectile state with momentum $\vec{q}$ and spin $\sigma$. In 
terms of pseudo-particle operators the physical meaning of the various terms of the Hamiltonian
is very transparent. For instance, the fourth line of~\eqref{ANM} denotes Auger neutralization, 
where the projectile transfers from the ion configuration to the groundstate configuration 
while creating an electron in state $\vert\vec{k}^\prime\sigma\rangle$ 
and destroying two metal electrons, one in state $\vert\vec{k}_1\downarrow\rangle$ and one in 
state $\vert\vec{k}_2\sigma\rangle$.

The energies $\varepsilon_{g}(t)$, $\varepsilon^*_{s}(t)$, $\varepsilon^*_{t}(t)$, and
$\varepsilon_{\vec{q}\sigma}(t)$ of the projectile's electronic configurations are time-dependent 
because of polarization effects which depend on the projectile's distance to the surface and 
thus its instantaneous position in front of the surface. We approximate this by an image 
shift as explained in~\cite{MBF12a,PBF15a}. The time-dependencies of the matrix elements
$V^{\rm SET}_{\vec{k}\sigma}(t)$ for single electron transfer, 
$V^{\rm AN}_{{\vec{k}_1}{\vec{k}_2}{\vec{k}^\prime\sigma\downarrow}}(t)$ for Auger neutralization, 
$V^{\rm DAD}_{\vec{k}\vec{k}\prime\sigma}(t)$ for direct Auger de-excitation, and
$V^{\rm IAD}_{\vec{k}\vec{q}\sigma}(t)$ for indirect Auger de-excitation arise from the overlap of 
projectile and target wave functions which also depend on the separation of the projectile and
the target. In our models the target is approximated by a three-dimensional potential step, 
its wave functions can thus be obtained analytically. For the bound states of the helium projectile 
we take hydrogen wave functions with charges adjusted to reproduce the ionization levels of helium. 
The projectile's continuum states are approximated by plane waves. Explicit expressions for the 
matrix elements will be given elsewhere~\cite{PBF17b}. 

Subjecting model \eqref{ANM} to a quantum kinetic treatment along the lines initially given 
by Langreth and coworkers~\cite{LN91,SLN94a,SLN94b}, a linear set of ordinary first order 
differential equations can be derived for the probabilities with which the projectile's 
electronic configurations occur in the course of the collision. In our case, we obtain
equations for the occurrence probabilities of the ion, the groundstate, and the two metastable 
states of the helium projectile. The derivation consists of three main steps. First, two-time Dyson 
equations for the projectile Green functions are set up. Second, the selfenergies are calculated 
in the non-crossing approximation and, third, the fact is utilized that the selfenergies are 
peaked around the time-diagonal enabling a saddle-point approximation to extract from the Dyson 
equations rate equations for the occurrence probabilities $n_i(t), n_t(t), n_s(t)$, and 
$n_g(t)$ for, respectively, the ion, the two metastable states, and the groundstate of the 
projectile. Following this reasoning which we also exerted in 
\cite{MBF12b,PBF15b,PBF15a} we get 
%\onecolumn

\begin{strip}
\begin{align}
\frac{d}{dt} \begin{pmatrix} 
n_{\rm i} \vspace{1mm}\\
n_{\rm t} \vspace{1mm}\\
n_{\rm s} \vspace{1mm}\\
n_{\rm g}
\end{pmatrix}
= 
\begin{pmatrix} 
-[\Gamma_{\rm t}^<(t) + \Gamma_{\rm s}^<(t) + \Gamma_{\rm AN}^<(t)] & \Gamma_{\rm t}^>(t) & \Gamma_{\rm s}^>(t) & \,\,\,0\,\,\, \vspace{1mm}\\
\Gamma_{\rm t}^<(t)  & -[\Gamma_{\rm t}^>(t) + \Gamma_{\rm IAD\uparrow}^<(t)] & 0      & \,\,0\,\, \vspace{1mm}\\
\Gamma_{\rm s}^< (t) & 0  & -[\Gamma_{s}^>(t) + \Gamma_{\rm IAD\downarrow}^<(t) + \Gamma_{\rm DAD\downarrow}^<(t)]  & \,\,0\,\, \vspace{1mm}\\
\Gamma_{\rm AN}^<(t)  & \Gamma_{\rm IAD\uparrow}^<(t) & \Gamma_{\rm IAD\downarrow}^<(t) + \Gamma_{\rm DAD\downarrow}^<(t) & \,\,0\,\, 
\end{pmatrix} 
\cdot 
\begin{pmatrix}
n_{\rm i} \vspace{1mm}\\
n_{\rm t} \vspace{1mm}\\
n_{\rm s} \vspace{1mm}\\
n_{\rm g} 
\end{pmatrix}  
\label{RateEq}
\end{align}
\end{strip}

%\twocolumn
\noindent
which can be solved numerically. Explicit expressions for the time-dependent rates 
$\Gamma_{...}^\gtrless(t)$ will be given elsewhere~\cite{PBF17b}. Notice, the entries in 
each column of the matrix sum up to zero. This is a consequence of the fact that the probability 
of finding the projectile at any particular instant of time in any one of its electronic 
configurations has to be one. Hence, $n_i(t)+n_t(t)+n_s(t)+n_g(t)=1$ implying
\begin{align}
\frac{d}{dt}[n_i(t)+n_t(t)+n_s(t)+n_g(t)]=0~.
\end{align}

\begin{figure}[t]
  \centering
  \includegraphics[width=0.99\linewidth]{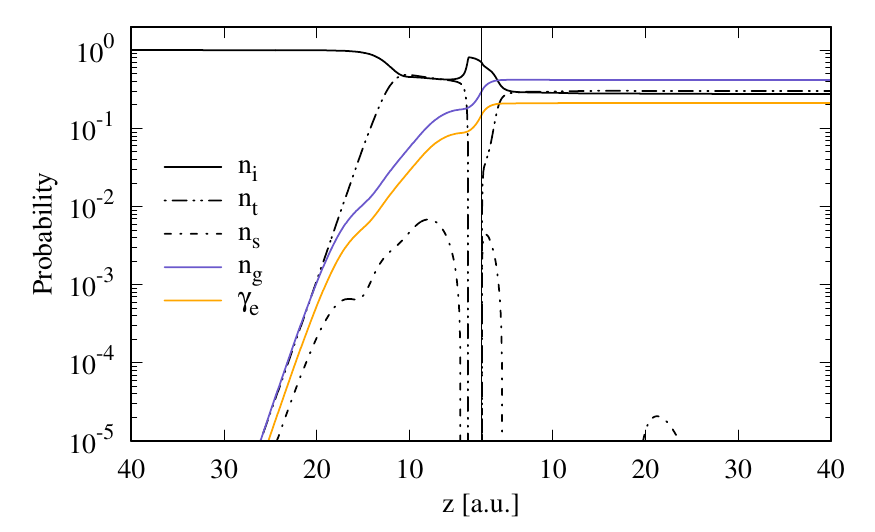}
  \caption{(Color online) Instantaneous occurrence probabilities $n_i, n_{t}, 
  n_{s}$, and $n_g$ of the electronic configurations of the helium projectile and 
  instantaneous electron emission probability $\gamma_e$ obtained from \eqref{RateEq}. 
  The projectile hits the tungsten surface as an ion with a kinetic 
  energy $E_{\rm kin}=200$ eV and an incident angle with respect to the surface of 
  $\varphi=2^\circ$. The turning point $z_{\rm tp}=2.27$ and is indicated by the thin 
  vertical line. Hence, the left (right) part of the plot denotes the incoming (outgoing) 
  branch of the collision trajectory. At the end of the collision, the projectile is 
  either still in its ionic state (black solid line), in its triplet metastable state
  (dot-dot-dashed line), or in its groundstate (blue solid line). The singlet metastable 
  state (dot-dashed line) occurs only temporarily. Its occurrence probability at the end 
  of the collision is vanishingly small. The secondary electron emission coefficient 
  is the value of $\gamma_e$ (orange solid line) at the end of the outgoing branch. It is 
  in good agreement with experimental data~\cite{MHB93}.  
  }
  \label{HeNeutralOcc}
\end{figure}

Representative data for the instantaneous occurrence probabilities $n_i(t), n_s(t), 
n_t(t)$ and $n_g(t)$ as well as the instantaneous secondary electron emission probability 
$\gamma_e(t)=n_g(t)/2$, assuming a fifty-fifty chance for the excited electron to leave
the solid, obtained from the numerical solution of \eqref{RateEq} are plotted in 
Figure~\ref{HeNeutralOcc} for a ${\rm He}^+(1^2{\rm S}_{1/2})$ ion hitting a tungsten 
surface with $E_{\rm kin}=200\,{\rm eV}$ under an angle of incident of $2^{\circ}$ 
measured from the surface. The parameter setting is not appropriate for a 
${\rm He}^+(1^2{\rm S}_{1/2})$ ion scattering off a plasma wall where the incident
would be normal and not grazing because of the sheath potential. However, the 
ion beam scattering experiments from which we obtained data for secondary electron 
emission coefficients have been performed for grazing incident~\cite{MHB93}. To 
compare our results with measured data we hence use this scattering geometry. 
Data for normal incident will be presented elsewhere~\cite{PBF17b}. For the 
situation depicted in Fig.~\ref{HeNeutralOcc} the helium projectile thus starts as 
an ion at a distance $z=40\,a_{\rm B}$ ($a_{\rm B}$ is the Bohr radius), moves along 
a grazing trajectory towards the turning point $z_{\rm tp}=2.27\, a_{\rm B}$, 
where it is specularly reflected to move back to the distance $z=40~a_{\rm B}$. After 
completion of the collision the projectile finds itself neutralized to the groundstate
${\rm He}(1^1{\rm S}_0)$ with probability $n_g(t_{\rm max})\approx 0.4$ and to the triplet 
metastable configuration ${\rm He}^*(2^3{\rm S}_1)$ with probability 
$n_t(t_{\rm max})\approx 0.3$. It thus survives the surface collision as an ion with probability 
$n_i(t_{\rm max})\approx 0.3$. 

The probability for emitting a secondary electron in the course of the collision is 
$\gamma_e(t_{\rm max})\approx 0.2$ which agrees surprisingly well with experimental 
data by M\"uller and coworkers~\cite{MHB93}. Despite the simplicity of the model and the 
uncertainties arising from the Gadzuk construction~\cite{Gadzuk67a,Gadzuk67b} we use for 
obtaining the matrix elements of the Hamiltonian \eqref{ANM}, we nevertheless obtain very 
good results. Even the spectrum of the emitted electron (not shown) turns out to be in good 
agreement with the measured spectrum. The result for the helium ion is thus rather encouraging. 
It is of course not a hard test of the viability of the semiempirical approach. For that we 
should at least be also able to reproduce experimental data~\cite{MHB93} for inert gas 
ions other than ${\rm He}^+(1^2{\rm S}_{1/2})$. Work in this direction is in progress. It 
requires the incorporation of $p-$orbitals and hence the renewed computation of all matrix 
elements and transition rates.

\section{Electric double layer}
\label{EDL}

Treating electron absorption/backscattering and electron extraction as elementary surface 
collision processes giving rise to a few surface parameters fed into the kinetic modeling of the 
plasma is a viable approach as long as the microphysics responsible for these processes is irrelevant 
on the scale of the plasma sheath. If this is however not the case, for instance, because of the 
miniaturization of the discharges, or if the physical system of interest consists of a plasma and a 
solid component which cannot be separated and have thus to be treated on an equal footing, as it is 
the case for the plasma bipolar junction transistor~\cite{WTE10}, or if one is simply interested per 
se in how the electronic non-equilibrium of the plasma is transferred to the electronic states of the 
solid, one has to employ kinetic equations not only for the plasma but also for the solid and to merge 
them by suitable matching conditions. Improving our earlier work~\cite{HBF12a} on the distribution of 
surplus electrons accumulated from the plasma inside a plasma-facing dielectric we recently set up a 
general kinetic framework~\cite{BF17b} for studying the charge transport across a dielectric plasma 
interface, that is, its electronic response, selfconsistently with the charge transport and relaxation 
on both sides of the interface. 

Our first attempt~\cite{HBF12a} of treating the plasma-induced electric double layer at a dielectric
plasma interface was based on a thermodynamic principle~\cite{TD85} and a graded 
potential~\cite{Stern78,SS84} interpolating between the potential of the sheath and the potential inside 
the solid. Due to the graded potential we could treat dielectrics with positive and negative electron 
affinities. The assumption of thermodynamic equilibrium of the surplus charges with the lattice of the 
solid, however, is too restrictive. It applies only to a subgroup of transferred electrons. In the new 
approach~\cite{BF17b} we overcome this limitation. The concept of an electron surface layer employed 
in~\cite{HBF12a} to describe the solid-based negative part of the electric double layer enabled us 
however to determine in which states the electrons accumulated from the plasma reside: For dielectrics 
with negative electron affinity they are confined in image states in front of the surface, forming thus 
a two-dimensional electron film as envisaged by Emeleus and Coulter~\cite{EC87,EC88} in their attempt
to go beyond the perfect absorber model for plasma walls, while for electro-positive dielectrics they 
populate the surface's conduction band, giving thus rise to a wide space charge region.

\begin{figure}[t]
  \centering
  \includegraphics[width=0.99\linewidth]{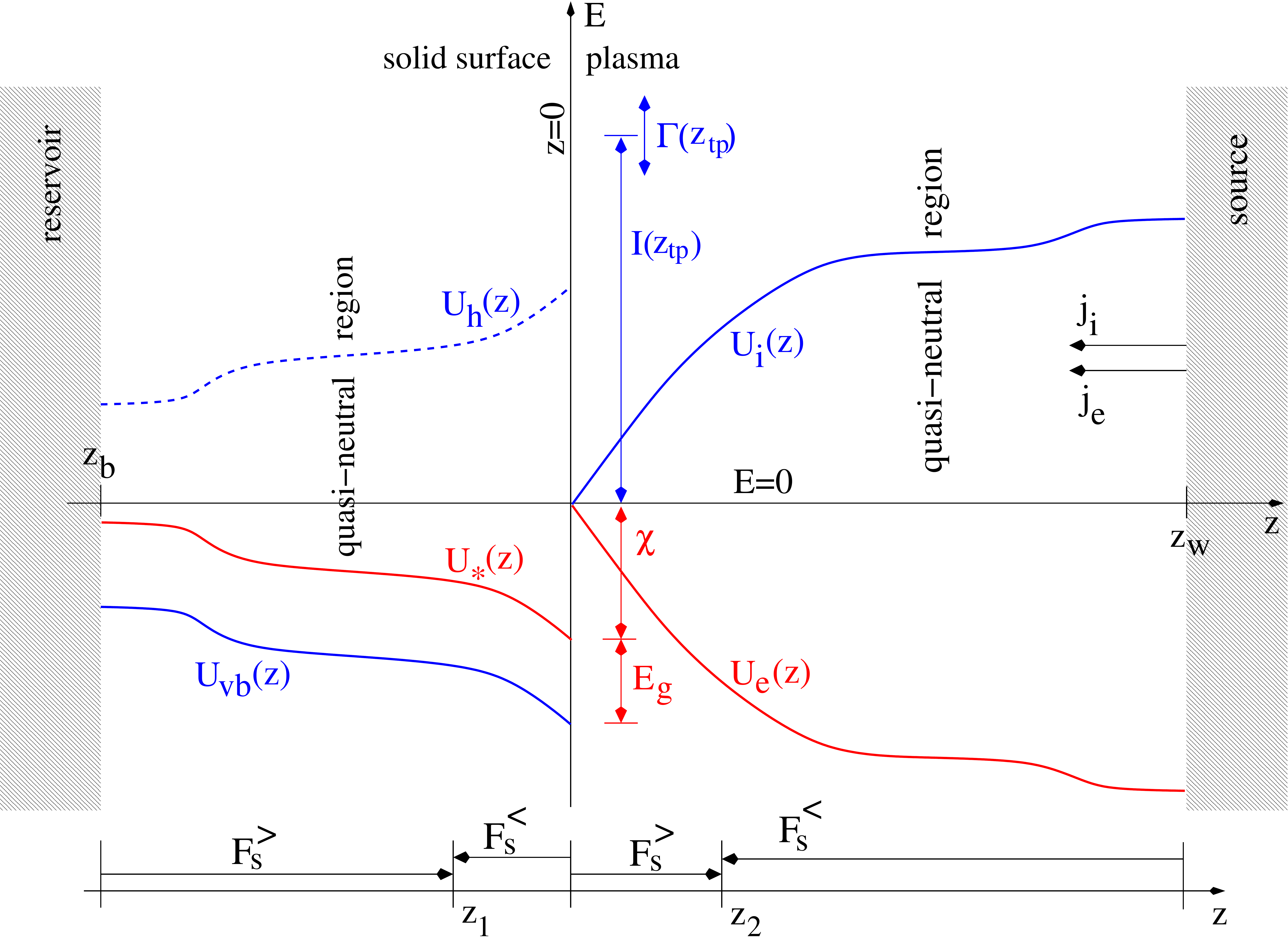}
  \caption{(Color online) Interface model for
        an electric double layer with negative space charge inside
        the solid and positive space charge in front of it~\cite{BF17b}. Shown are the edges of
        the conduction ($U_*$) and valence bands ($U_{\rm vb}$), the edge for the motion
        of valence band holes ($U_{\rm h}$), the potential energies for electrons and ions
        on the plasma side, and the energetic range, specified by the ion's ionization energy
        $I$ and its broadening $\Gamma$, in which hole injection occurs due
        to the neutralization of ions at the interface. Source, reservoir, and quasi-neutral
        regions are indicated as they will arise in the course of the calculation.}
  \label{EDLcartoon}
\end{figure}

The fundamental set of equations describing the electronic response of a plasma-facing
dielectric solid is given in~\cite{BF17b}. It consists of the Poisson equation
\begin{align}
\frac{d}{dz}\varepsilon(z)\frac{d}{dz} U_c(z) = 8\pi\rho(z) =
                                                8\pi\big[\rho_w(z)\theta(-z)-\rho_p(z)\theta(z)\big]
\label{PE}
\end{align}
and two sets of spatially separated Boltzmann equations, one for the electrons and ions in the plasma 
and one for the conduction band electrons and valence band holes in the solid. Defining a species
index $s=e,i,*,h$ to denote electrons, ions, conduction band electrons, and valence band holes,
respectively, the Boltzmann equations for the quasi-stationary distribution functions 
$F_s^\gtrless(z,E,\vec{K})$, for left and right moving particles (see Figure~\ref{EDLcartoon}), can be 
written as
\begin{equation}
\left[\pm v_s(z,E,\vec{K})\frac{\partial}{\partial z}+\gamma_s[F_{s^\prime}^{\gtrless}]\right]
F_s^{\gtrless}(z,E,\vec{K})=\Phi_s[F_{s^\prime}^{\gtrless}]
\label{BTE}
\end{equation}
with
\begin{equation}
v_s(z,E,\vec{K}) = 2\bigg(\frac{m_e}{m_s}[E-U_s(z)] - (\frac{m_e}{m_s}\vec{K})^2 \bigg)^{1/2}~
\label{velocity}
\end{equation}
the velocity of the particles normal to the (planar) interface at $z=0$ and $z$, $E$, and
$\vec{K}$ the distance from the interface, the total energy, and the lateral momentum. 
The potential energies entering \eqref{velocity} are given by 
\begin{align}
U_*(z) &=-U_c(z)- \chi ~,~~U_e(z)=-U_c(z)~,\\
U_h(z) &=U_c(z)+E_g+\chi ~,~~U_i(z)=U_c(z)~
\end{align}
with $U_c(z)$ the solution of~\eqref{PE}, $E_g$ the band gap, and $\chi>0$ the electron 
affinity. Energies and lengths are again measured in Rydbergs and Bohr radii.
\begin{figure}[t]
  \centering
  \includegraphics[width=0.99\linewidth]{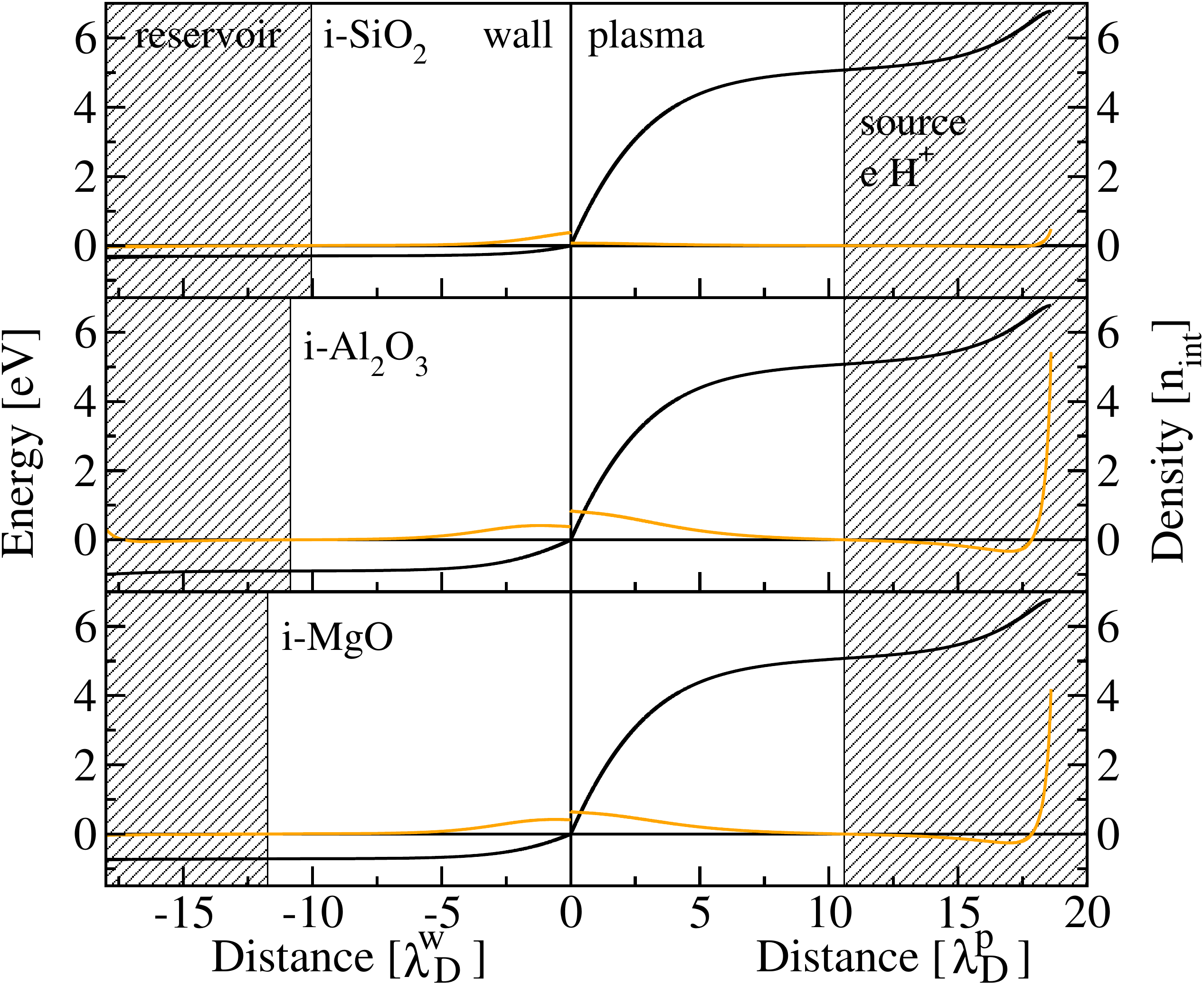}
  \caption{(Color online) Potential energy (black lines) and density (orange
  lines) profiles defined by $\rho^t_w(z)=[n^t_*(z)-n^t_h(z)]\theta(-z)$ and
  $\rho_p(z)=[n_i(z)-n_e(z)]\theta(z)$ for i-\SiOTwo~\cite{BF17b}, i-\AlTwoOThree, and
  i-\MgO\, surfaces in contact with a hydrogen plasma with $k_BT_e=10k_BT_i=2\, {\rm eV}$.
  Only the thermalized part of the charge density inside the wall, $\rho^t_w(z)$,
  is shown which balances $\rho_p(z)$ on the plasma side. The interface is assumed to
  be perfectly absorbing and collisionless on both sides. Grey regions on the left and
  right indicate respectively the reservoir and the source required for implementing
  the physical boundary conditions for the electric double layer. Relevant for
  the double layer is only the region between the two thin vertical lines
  indicating the positions of the inflection points $z_1<0$ and $z_p>0$ in units
  of the Debye screening lengths $\lambda_D^w$ and $\lambda_D^p$, respectively.
  The band bending is given by $U_1=U_c(z_1)$ while the sheath potential
  becomes $U_p=U_c(z_p)$. Material parameters used for the calculations are
  summarized in Table~\ref{MaterialParameters}.}
  \label{EDLprofiles}
\end{figure}

The functions $\gamma_s[F_{s^\prime}^{\gtrless}]$ and $\Phi_s[F_{s^\prime}^{\gtrless}]$ denote,
respectively, the rates for out-scattering and the in-scattering collision integrals. The set
of Boltzmann equations has to be augmented by matching conditions for the distribution
functions. For electrons they can be derived quantum-mechanically using a technique developed
for solid surfaces~\cite{Falkovsky83} and solid-solid heterostructures~\cite{DLP95,Schroeder92}. 
So far the technique has been applied to match bulk electron states across interfaces. In 
principle it should be however possible to generalize the approach to take surface states into 
account as well. The matching of the ion and hole distribution functions on the other hand has 
to be based on a physical model for hole injection due to, for instance, resonant neutralization 
of ions at the interface. Neutralization due to Auger processes leading in addition to the emission 
of a secondary electron as discussed in Section \ref{Extraction} could be also treated but we 
have not done so yet. 

For a dielectric plasma interface without surface states and for resonant ion neutralization only 
the matching conditions for the distribution functions become~\cite{BF17b}
\begin{align}
F_{e,*}^{>,<}(0,E,\vec{K}) &= R(E,\vec{K}) F_{e,*}^{<,>}(0,E,\vec{K}) \nonumber\\
                   &+ [1-R(E,\vec{K})]F_{*,e}^{>,<}(0,E,\vec{K})~
%F_*^<(0^-,E,\vec{K}) &= R(E,\vec{K}) F_*^>(0^-,E,\vec{K})
%                   + [1-R(E,\vec{K})]F_e^<(0^+,E,\vec{K})
\end{align}
with $E>0$ and 
\begin{align}
F_h^<(0,E,\vec{K}) &= F_h^>(0,E,\vec{K}) + \alpha S_h^<(E,\vec{K})~,~ \\
F_i^>(0,E,\vec{K}) &= (1-\alpha)F_i^<(0,E,\vec{K})
\label{Fi_gtr_match}
\end{align}
with $E>E_g+\chi$, where $R(E,\vec{K})$ is the quantum mechanical reflection coefficient 
for electrons due to the surface potential, $\alpha$ is the ion neutralization probability, and
$S_h^<(E,\vec{K})$ is a function specifying hole injection into the valence band of the
dielectric due to the neutralization of an ion at the interface. The particular form of
\begin{figure}[t]
  \centering
  \includegraphics[width=0.99\linewidth]{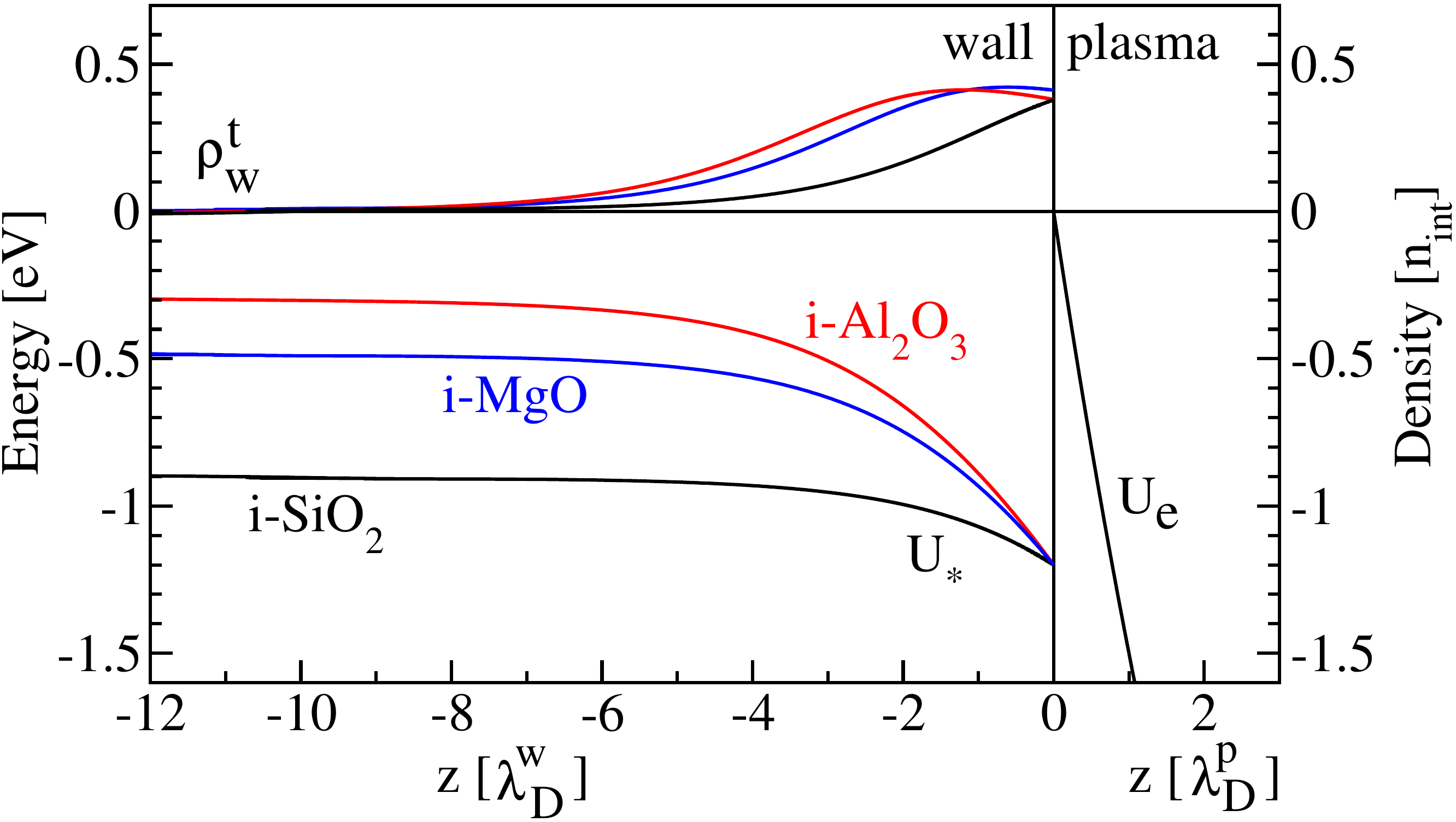}
  \caption{(Color online) Conduction band edges $U_*(z)$  and
  density profiles $\rho_w^t(z)$ for i-\SiOTwo, i-\AlTwoOThree,
  and i-\MgO\, surfaces in contact with an hydrogen plasma with
  $k_BT_e=10k_BT_i=2\, {\rm eV}$. Material parameters can be
  found in Table~\ref{MaterialParameters}. The interface is perfectly
  absorbing and collisionless on both sides.}
  \label{EDLbending}
\end{figure}
$S_h^<(E,\vec{K})$ depends on the neutralization process. In case the neutralization
induces also secondary electron emission, the matching condition for the electron
distribution functions has to be augmented by a function $S_e^>(E,\vec{K})$. The 
functions $S_h^<(E,\vec{K})$ and $S_e^>(E,\vec{K})$ have thus to account for the processes 
discussed in Section~\ref{Extraction}. 

Together with the Poisson equation~\eqref{PE} for the electric potential energy $U_c(z)$, 
the matching conditions for $U_c(z)$ at $z=0$, and 
\begin{eqnarray}
n_s(z)=\int\frac{dE d^2K}{(2\pi)^3} \frac{F^>_s(z,E,\vec{K})+F^<_s(z,E,\vec{K})}{v_s(z,E,\vec{K})}~,
\label{Density}
\end{eqnarray}
for the charge densities, Eqs.~\eqref{BTE}--\eqref{Fi_gtr_match} form a closed set of equations 
for the distribution functions provided they are supplemented by boundary conditions far away 
from the interface, that is, at $z=z_b<0$ and $z=z_w>0$ (see Figure~\ref{EDLcartoon}).
For a floating plasma interface the boundary condition at $z=z_w$ is the quasi-neutral bulk
plasma, acting as a source for ions and electrons, whereas at $z=z_b$ an intrinsic or extrinsic
dielectric has to be established as a quasi-neutral reservoir for conduction band electrons
and valence band holes. The physical picture arising for a floating interface is illustrated
in Figure~\ref{EDLcartoon} anticipating an electron-rich (electron-depleted)
surface (plasma sheath). Electrically contacted plasma interfaces (electrodes) require other
boundary conditions and have to be modelled differently.

The approach presented so far is quite general. To obtain results scattering processes
have to be specified. On the plasma side the most important scattering process is 
charge exchange scattering for ions whereas inside the solid intraband electron-phonon
scattering and interband electron-hole recombination should be included. Once the 
collision integrals associated with these processes are worked out the set of 
Boltzmann equations has to be solved numerically together with the Poisson equation,
the matching conditions, and the boundary conditions. This is quite demanding and part 
of our future activities in this field. 

To obtain first estimates for the electric 
double layer forming due to the electronic response of the interface, we considered in 
\cite{BF17b} the case of a collisionless, perfectly absorbing interface. In this 
particular case the lateral momentum $\vec{K}$ can be eliminated. It is then possible
to formulate the whole approach in terms of 
\begin{eqnarray}
F_s(z,k)=\int\frac{d^2K}{(2\pi)^2} F_s(z,\vec{K},k) ~,
\end{eqnarray}
which satisfy 
\begin{equation}
\pm v_s(z,E)\frac{\partial}{\partial z}F_s^{\gtrless}(z,E)=0
\label{BTECollLess}
\end{equation}
with 
\begin{eqnarray}
v_s(z,E) = 2\bigg(\frac{m_e}{m_s}[E-U_s(z)]\bigg)^{1/2}.
\end{eqnarray}
The Boltzmann equations become thus ordinary differential equations which can 
be solved, as explained in~\cite{BF17b}, by a straightforward trajectory analysis 
taking at $z=0$ the perfect absorber matching conditions and at $z=z_{b,w}$ the 
boundary conditions into account specified in the previous paragraph. For details 
see ~\cite{BF17b}. It turns out that in this particular case the densities $n_s(z)$ 
are only functions of $U_c(z)$ and not of $z$ explicitly. Hence, the Poisson equation
can be integrated once analytically and a second time numerically to obtain $U_c(z)$. 

Numerical results for this simple case are shown in Figures~\ref{EDLprofiles} and 
\ref{EDLbending} using a hydrogen plasma in contact with intrinsic \SiOTwo, 
\AlTwoOThree, and \MgO\, surfaces as examples. The material parameters are given 
in Table~\ref{MaterialParameters}. Notice in particular the plasma-induced band bending 
inside the dielectrics. As in other electronic devices, it would control together with 
the band off-set $\chi$ the electric current in case the interface was wired up to an 
external circuit. To make the approach selfconsistent, it is necessary to split the 
charge density inside the solid into a thermalized (trapped) and a non-thermalized (non-trapped) 
part: $\rho_w(z)=\rho_w^t(z)+\rho_w^j(z)$. Roughly speaking, $\rho_w^t(z)$ is a consequence
of the boundary condition at $z=z_b$ while $\rho_w^j(z)$ is a consequence of the 
matching condition at the interface. In effect this leads to a two-species scenario 
and a recombination condition limiting in conjunction with the boundary conditions 
the influx $j=j_e=j_i$ of electrons and ions to the interface. Numerical data for the 
influx are given with other quantities of interest in Table~\ref{TableForData}. Notice, 
even in the simply case considered here--the collisionless, perfectly absorbing 
plasma interface--the influx $j$ and the total wall charge 
$N_{\rm w}=\int_{z_1}^0 dz \rho_w^t(z)$ establishing quasi-stationarity depend on the
material. The reason is the selfconsistent adjustment of the plasma source to the 
environment inside the solid characterized here by the intrinsic charge density 
$n_{\rm int}$ (given in Table~\ref{TableForData}), the energy gap $E_g$ , and the 
temperatures of the thermalized holes and electrons, $T_h$ and $T_*$. In a realistic model, 
where collisions are included, in particular, interband collisions annihilating electron-hole 
pairs, the adjustment of the plasma source to the conditions inside the wall will become 
even more pertinent.

\begin{figure}[t]
        \begin{minipage}{0.35\linewidth}
        \centering
        \includegraphics[width=0.98\textwidth]{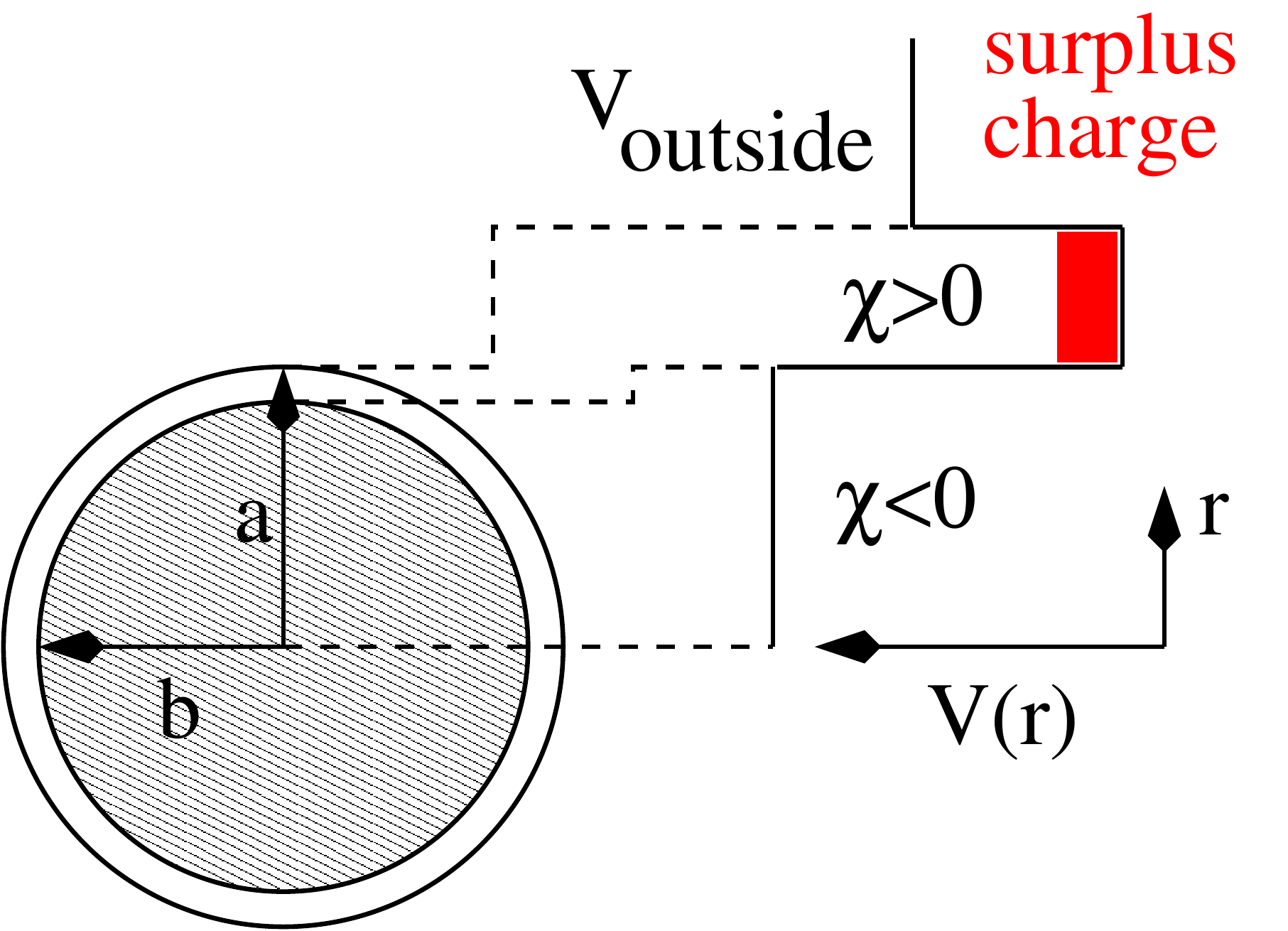}
        \end{minipage}\begin{minipage}{0.65\linewidth}
        \centering
        \includegraphics[width=0.98\textwidth]{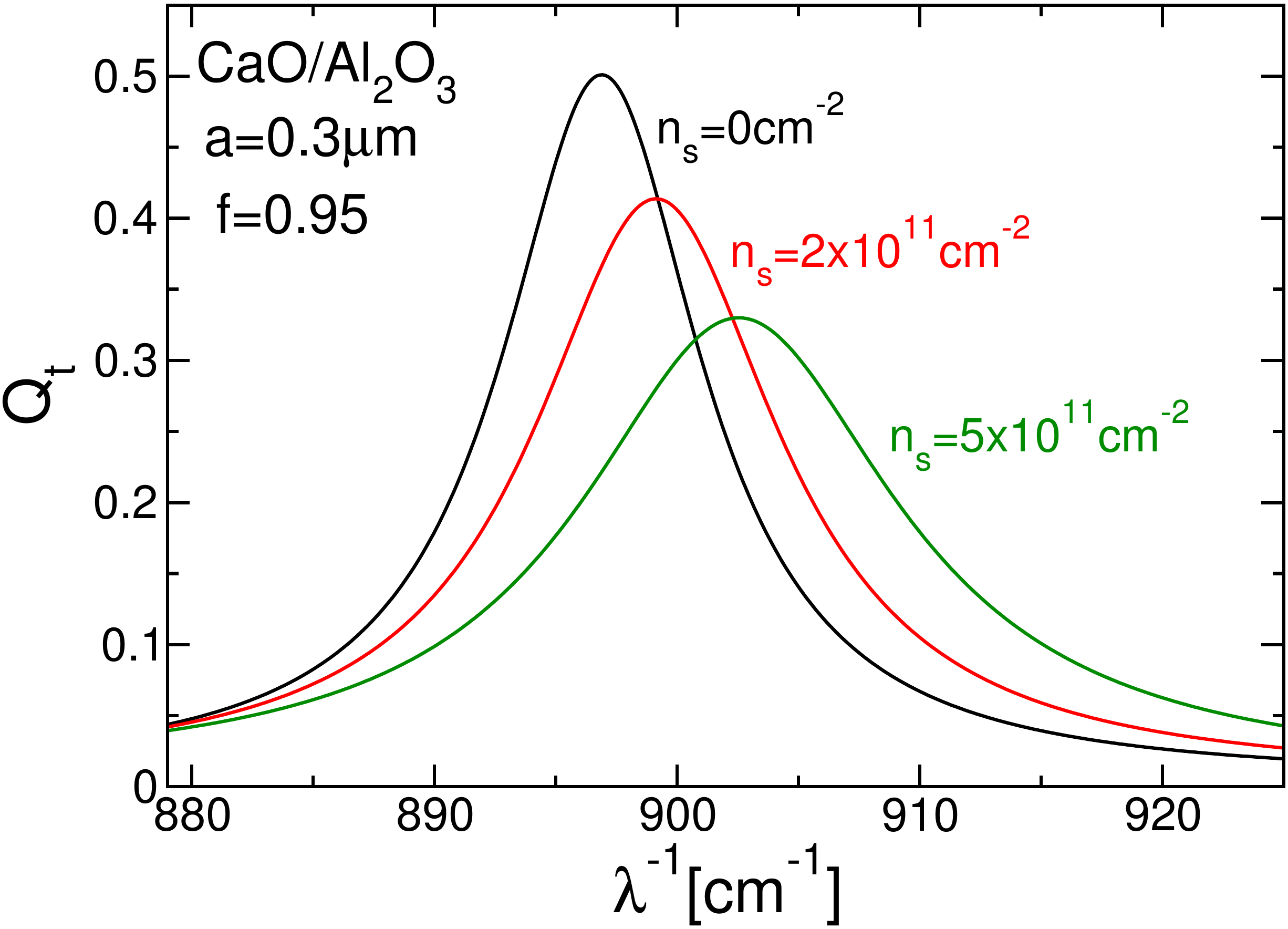}
        \end{minipage}
        \caption{(Color online) Geometry and profile of the conduction band edges for a
                 ${\rm CaO}/{\rm Al}_2{\rm O}_3$ core-shell particle confining surplus electrons
                 accumulated from the plasma in a narrow shell of thickness $d=(1-f)a$ with $f=b/a$
                 the filling factor. The antibonding resonance arising from the surface modes
                 at the two interfaces shifts with charge density (data adapted from~\cite{THB14}.)}
        \label{MieSummary}
\end{figure}

\section{Spectroscopy of the wall charge}
\label{EELS}

In an effort to establish a method for measuring the charge of dust grains in a plasma 
optically we investigated Mie scattering by charged dielectric particles. Initially 
we considered homogeneous particles with negative and positive electron 
affinity~\cite{HBF13,HBF12b} but later we also investigated core-shell particles, 
where an electro-negative core is coated by an electro-positive film~\cite{THB14}. 
The idea was based on the observation that the particle charge modifies either the
boundary condition for the electromagnetic field (particles with negative electron
affinity) or the bulk dielectric function (particles with positive electron affinity).
In either case the optical extinction $Q_t$ depends on the charge of the particle which 
should thus be visible in the Mie signal calculated and measured by standard 
procedures~\cite{BH83}.

We found light scattering by particles made out of dielectric materials featuring
transverse optical phonons, for instance, MgO, CaO, ${\rm Al}_2{\rm O}_3$ to be very charge
sensitive in the vicinity of anomalous optical resonances~\cite{HBF13,HBF12b},
defined by $\varepsilon^\prime (\omega) < 0$ and $\varepsilon^{\prime\prime}(\omega) \ll 1$,
where $\varepsilon(\omega)=\varepsilon^\prime(\omega)+i\varepsilon^{\prime\prime}(\omega)$ is
the complex dielectric function of the material~\cite{TL06,Tribelsky11}. The resonances are in 
the infrared and shift with increasing charge towards higher wave numbers. Confining the charge 
to a narrow electro-positive shell around an electro-negative core increases the charge sensitivity
of the shift~\cite{THB14}. In  Figure~\ref{MieSummary} the geometry of such a composite 
particle is shown together with results for a CaO/${\rm Al}_2{\rm O}_3$ core-shell 
particle. It turns out~\cite{THB16} that the antibonding resonance, arising from the hybridization 
of the two surface modes~\cite{PRH03,PS03} localized, respectively, at the core-shell and the 
shell-vacuum interface (that is, in plasma applications, the shell-plasma interface) is most 
charge-sensitive. Core-shell particles with this morphology could thus be used as electric probes 
with optical read-out.

\begin{figure*}[t]
        \centering
        \begin{minipage}{0.2\linewidth}
        \centering
        \includegraphics[width=0.90\textwidth]{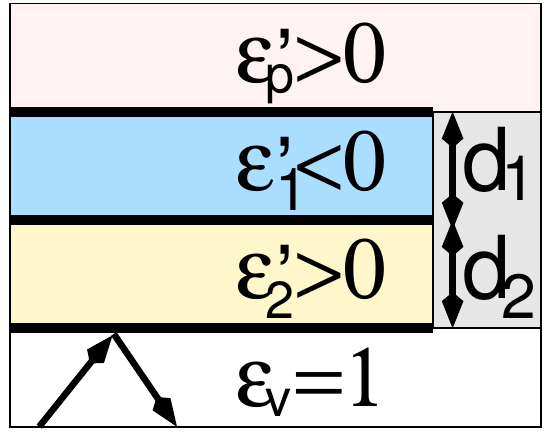}
        \end{minipage}\begin{minipage}{0.35\linewidth}
        \centering
        \includegraphics[width=0.93\textwidth]{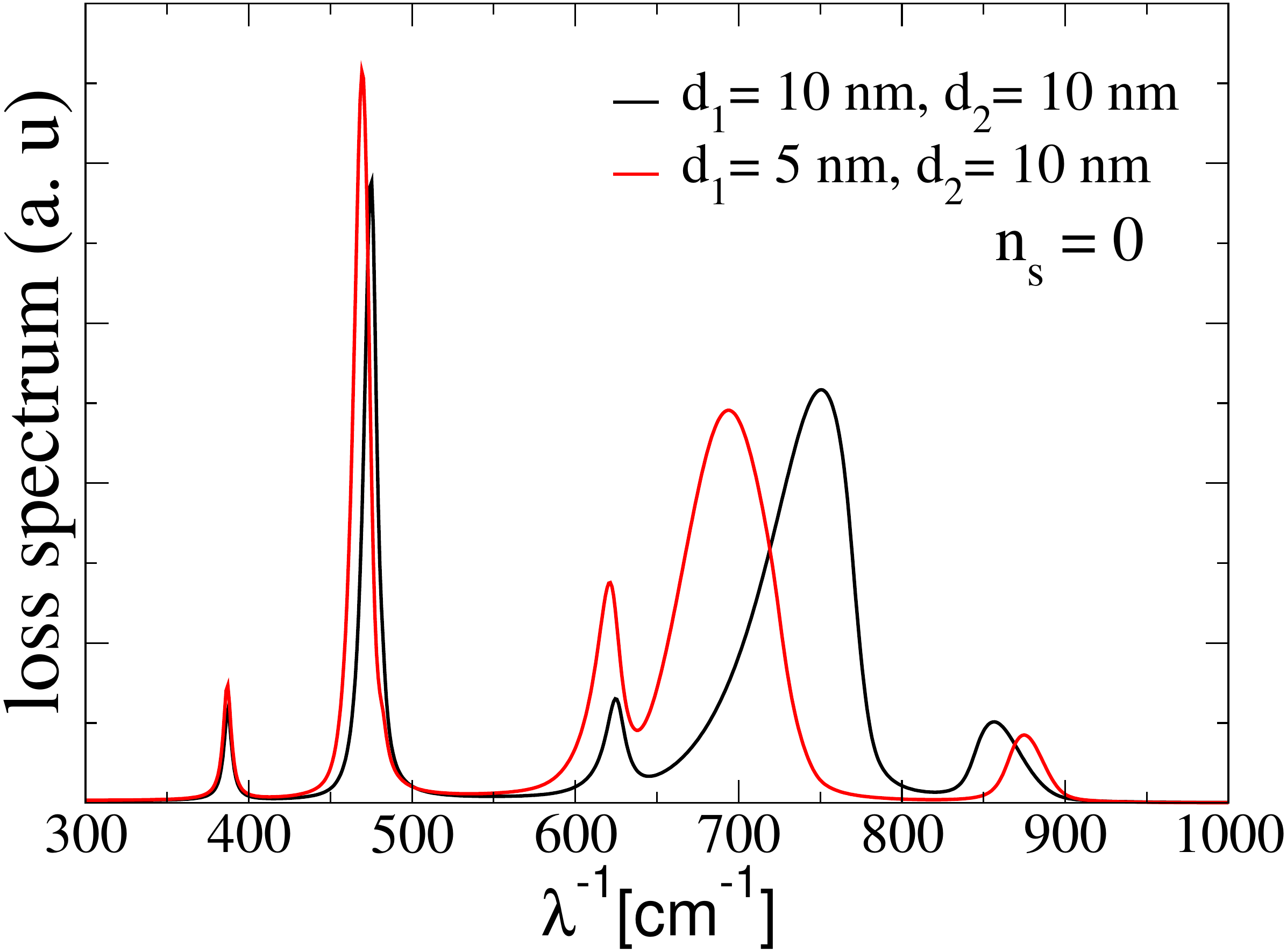}
        \end{minipage}\begin{minipage}{0.25\linewidth}
        \centering
        \includegraphics[width=0.98\textwidth]{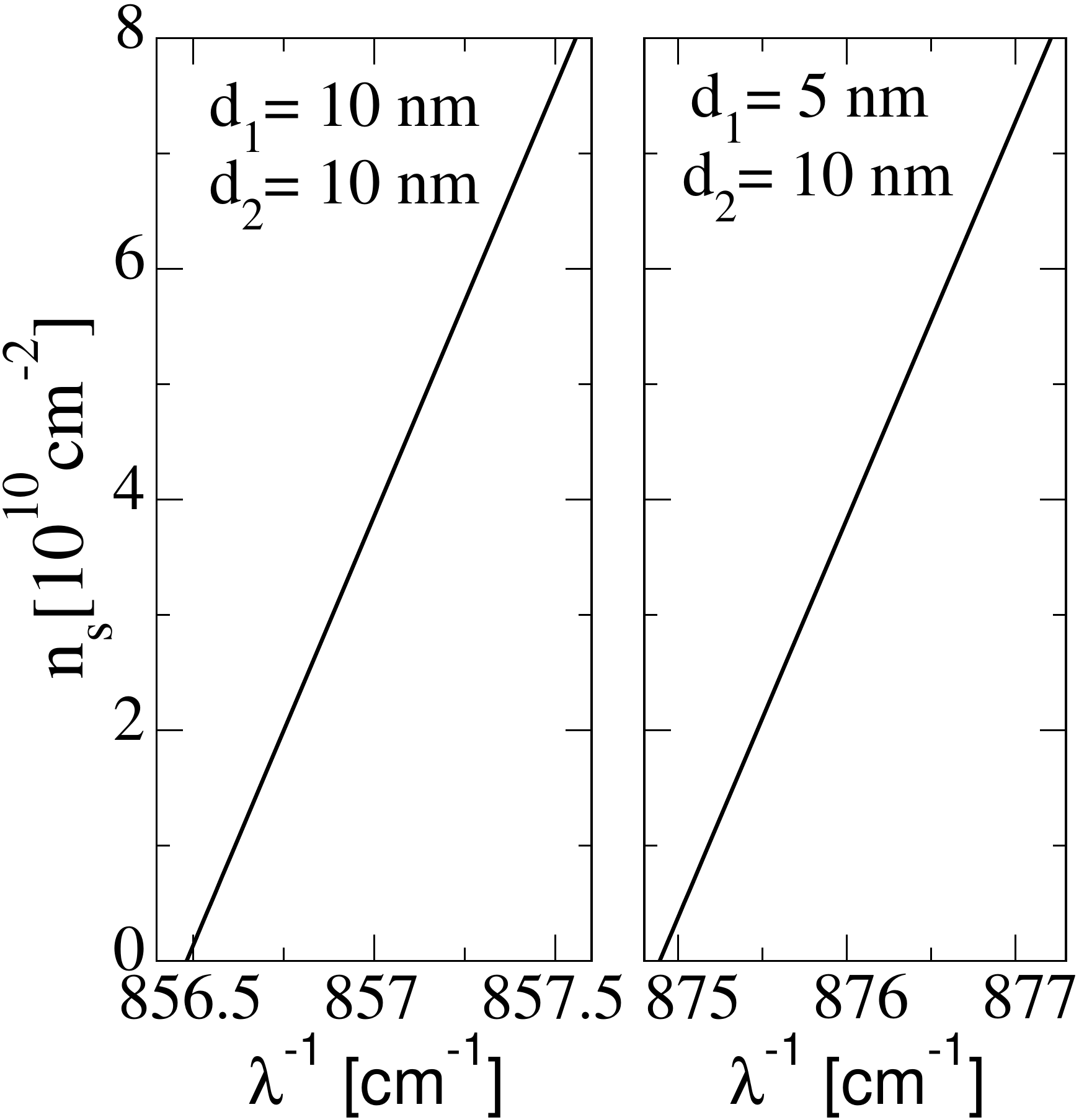}
        \end{minipage}
        \caption{(Color online) Results for the electron energy loss spectrum of a stack 
         of two dielectric layers in contact with a plasma providing the 
         dielectric mismatch at the uppermost interface and the surplus charges inside 
         the plasma-facing layer (although the charge transfer from 
         the plasma to the layer is not explicitly modelled). The 
         charge-carrying layer has thickness $d_1$ while the substrate has thickness
         $d_2$. The most charge-sensitive feature of the loss spectrum is the peak in
         the range $850-870\,{\rm cm}^{-1}$ which is the antibonding resonance arising
         from the modes at the dielectric/dielectric and the dielectric/plasma
         interface. Its shift with charge density is plotted
         on the right for two different thicknesses $d_1$ of the charge-carrying layer.}
        \label{EELSdata}
\end{figure*}

The idea of using engineered solid-state objects as electric probes can be also applied to planar
walls. Indeed, the surface modes responsible for the charge-sensitivity of Mie scattering by 
small particles become the surface-plasmon polaritons in the case of planar walls. Their dispersion and 
damping depends on the dielectric function of the wall which in turn is modified by the surplus
electrons accumulated from the plasma. Spectroscopic techniques sensitive to the surface-plasmon 
polariton should thus be capable to reveal information about the space charge inside the 
solid. Hence, using such a spectroscopy in plasma applications would provide a sensor for the 
physical properties of the wall charge. Particularly promising in this respect seems to be 
electron-energy loss spectroscopy~\cite{IbachMills82,LVL85} which demonstrated for free-standing
surfaces and solid heterostructures its feasibility for investigating buried space 
charge regions~\cite{MVM04,Lueth88,EM87}.

An exploratory theoretical study of small-angle inelastic electron scattering on a dielectric
heterostructure consisting of a charge-carrying layer of thickness $d_1$, in contact with a plasma,
and a substrate layer of thickness $d_2$ showed that the charge dependence of the effective
dielectric function governing the scattering of the electron is strong enough to produce 
measurable effects also in cases where the electron beam is applied to the substrate layer 
and not to the charge-carrying layer as shown in the left panel of Figure~\ref{EELSdata}.
Similar to Mie scattering by charged dielectric core-shell particles~\cite{THB14}
the antibonding resonance arising from the two surface modes at, respectively,
the dielectric/dielectric and the dielectric/plasma interface turns out to be particularly 
charge sensitive. The charge-dependent shift of its spectral position is shown on the right 
side of Figure~\ref{EELSdata}. It is the larger the thinner the charge-carrying layer, that is, 
the stronger the surplus electrons are confined.

To obtain the data shown in Figure~\ref{EELSdata} we assumed electrons with surface density $n_s$ to be 
homogeneously distributed in the charge-carrying layer and calculated, using $n_s$, $d_1$, and $d_2$ 
as parameters, the electron energy loss spectrum~\cite{LVL85}, 
\begin{align}
\frac{dS}{d\omega}=\int d^2 q_\parallel A(\vec{q}_\parallel,\omega)
                   {\rm Im}\bigg[\frac{-1}{\varepsilon_{\rm eff}(\vec{q}_\parallel,\omega)+1}\bigg]~,
\label{ElossFct1}
\end{align}
where the scattering kinetics is accounted for by the function 
\begin{align}
A(\vec{q}_\parallel,\omega)=\frac{4e^2}{\pi^2\hbar}
\frac{q_\parallel v_\perp^2}{[(\omega-\vec{q}_\parallel\cdot\vec{v}_\parallel)^2+q_\parallel^2 v_\perp^2]^2}’,
\end{align}
with $\vec{v}_\parallel$ and $v_\perp$ the lateral and perpendicular velocity of the electron 
with respect to the interface, and the energy loss is encoded into an effective dielectric
function 
\begin{align}
\varepsilon_{\rm eff}(\vec{q}_\parallel,\omega)=a_2-\frac{b_2^2}{a_2+a_1-b_1^2/(a_1+a_p)}
\end{align}
with $a_i=\varepsilon_i\coth(q_\parallel d_i)$ and $b_i=\varepsilon_i/\sinh(q_\parallel d_i)$ for 
$i=1, 2, p$ and $b_p=0$. Neglecting nonlocal effects, the surplus charges affect the effective 
dielectric function only through the dielectric function of the charge-carrying layer,
\begin{align}
\varepsilon_1(\omega) \rightarrow \tilde{\varepsilon}_1(\omega)=
\varepsilon_1(\omega)+4\pi i \sigma(\omega)/\omega~,
\end{align}
where $\sigma(\omega)$ is the phonon-limited conductivity of a gas of electrons with
volume density $n_b=n_s/d_1$ to be obtained in the same way as in~\cite{THB14}. 

The numerical results plotted in Figure~\ref{EELSdata}, obtained for a charge-carrying 
layer made out of \AlTwoOThree\ and a dissipationless substrate layer specified by 
$\varepsilon_2=2$, demonstrate that the charge-induced 
modification of the effective dielectric function can be strong enough to be detectable by 
a scattering geometry where the beam of electrons suffering energy loss is not in direct 
contact with the charge-carrying layer. This is important if one wants to employ 
electron energy loss spectroscopy as a probe for the charge build-up in the wall 
of a bounded plasma where the charged species and the electric fields in the plasma 
prevent the probing electron beam to be applied to the plasma-facing side of
the wall. To measure the charge distribution $\rho_w(z)$ accumulated in a plasma-facing 
solid structure can thus not be performed in the (standard) scattering geometry developed,
for instance, for investigating space charge layers at semiconducting 
surfaces~\cite{MVM04,Lueth88,EM87}. Our results demonstrate however that from-the-back 
diagnostics of $\rho_w(z)$ or $N_{\rm w}$, suitable for plasma applications, should be 
feasible.

\section{Concluding remarks}

In a still on-going effort we work towards a microscopic understanding of the electron 
kinetics across the plasma interface. The main motivation for our work, which we reviewed
and extended in this presentation in a colloquial manner--the mathematical details of which 
can be found in the original publications or in the work to be published--stems from the 
prospects of solid-based integrated microdischarges.
We expect the charge transfer between a solid and a plasma to be controllable by a judicious 
design of the plasma-facing structure opening up thereby routes to a new generation of 
opto-electronic plasma devices. Even if this expectation does not become reality conventional 
discharges also benefit from a better understanding of the charge transfer between the plasma and 
the solid because it leads to more realistic electron absorption/backscattering and secondary 
electron emission coefficients.

As far as electron absorption/backscattering and electron extraction due to atomic particles as 
elementary surface collision processes are concerned we presented general theoretical frameworks 
for calculating electron sticking/backscattering and secondary electron emission probabilities. 
In particular the invariant embedding approach used for investigating the interaction of a 
low-energy electron with an electro-positive dielectric surface has great potential. It can be 
extended to higher electron energies, above the impact ionization threshold of the dielectric, and 
also applied to metallic surfaces, if it is augmented by Coulomb-driven backscattering processes. 
The semiempirical multi-channel Anderson-Newns model we use for calculating secondary electron 
emission probabilities due to low-energy atom-surface collisions (potential ejection of electrons)
leads to very promising results for the projectile-target combinations 
we applied it to despite the uncertainties of the Gadzuk construction used for the calculation 
of the matrix elements. Further studies along this line, and comparison with results obtained 
from more refined treatments, are however necessary to demonstrate that the semiempirical model
produces indeed viable results for potential electron ejection due to charge-transferring 
atom-surface collisions. Kinetic ejection of electrons, occurring at impact energies on the 
order of 100~eV, has to be treated differently, because the projectile has then enough energy to 
enter the solid leading then to electron emission due to a cascade of collisions inside the solid.

We also presented the fundamental set of equations describing at the semiclassical level the 
selfconsistent electronic response of a plasma-facing solid. It consists of the Poisson equation 
for the electric potential energy and two sets of spatially separated Boltzmann equations, one 
for the electrons and ions inside the plasma and one for the conduction band electrons and valence 
band holes inside the solid. The solutions of the two sets of Boltzmann equations are 
connected by matching conditions. For electrons the matching conditions describe quantum-mechanical 
reflection and transmission due to the abrupt change in potential energy at the plasma 
interface whereas for ions and holes the matching conditions mimic hole injection (and thus 
electron extraction) due to resonant  
neutralization of ions. Secondary electron emission due to high energy electrons impacting the 
surface and/or Auger neutralization (de-excitation) of ions (metastable states) could be also included 
into the matching conditions. In an exploratory investigation we applied the approach to a perfectly
absorbing collisionless plasma interface and obtained first estimates for the plasma-induced 
band bending and the spatial charge distribution of the solid-based part of the electric 
double layer. Making the interface model more realistic by going beyond the perfect absorber 
assumption, accounting for collisions on both sides of and at the interface, and  
including the full band structure of the surface is of course necessary to produce reliable 
data. All this can be however done within the general framework presented above resulting 
in a kinetic model revealing--for the first time--the basic mechanisms coupling the charge 
production on the gaseous side of the interface with the charge losses inside the solid.

Based on our proposal of using spectroscopic means to determine the charge of a micron-size dust 
particle in a plasma we explored furthermore the possibility to employ electron-energy loss 
spectroscopy to probe the charge accumulated in a planar plasma-facing solid. Having experimental
access to the solid-based negative part of the electric double layer at a plasma interface 
will be crucial for developing a quantitative microscopic understanding of the charge transfer
across the interface. Our preliminary results indicate electron-energy loss spectroscopy to be 
indeed sensitive to the charge distribution inside plasma-facing solid structures. 
We expect it thus to be a key experimental technique for quantifying plasma-induced space charge
layers in solids.

\begin{acknowledgement}
Support from the Deutsche Forschungsgemeinschaft through Project B10 
of the second and third funding period of the Transregional Collaborative Research Center 
SFB/TRR 24 is greatly acknowledged. The contributions of R. L. Heinisch and J. Marbach 
in the second funding period are also greatly acknowledged.  
\end{acknowledgement}

\section*{Author contribution statement}
F. X. B. and H. F. outlined the scope and strategy of the calculations on which 
the work presented here is based; the computations of the electron sticking 
and backscattering coefficients as well as the calculations devoted to the 
electric double layer were performed by F. X. B.; M. P. and E. T. performed, 
respectively, the calculations concerning the neutralization of helium ions and the 
spectroscopy of the wall charge. The paper was written by F. X. B.; all authors contributed 
to the work presented here and approved the final text of the manuscript.

\bibliographystyle{epj}
\bibliography{ref}

\newpage

\begin{table*}[t]
%  \begin{tabular}{c|cccc|ccc|cccc}
  \begin{tabular}{|l||lll||lll||lll|}
  \hline
%                    &                     &         &          &                     &        &        &                          &                         &    \\
                    & $\lambda_D^w$       & $U_1$   & $U_b$    & $\lambda_D^p$       & $U_p$  & $U_w$  & $n_{\rm int}$            & $N_{\rm w}$          & $j$ \\
                    & ($10^{-3}{\rm cm}$) & (eV)    & (eV)     & $(10^{-3}{\rm cm}$) & (eV)   & (eV)   & $(10^{10}{\rm cm}^{-3}$) & $(10^{7}{\rm cm}^{-2}$) & ($10^{15}{\rm s}^{-1}{\rm cm}^{-2}$) \\\hline
  i-\SiOTwo           & 2.4                 & -0.3    & -0.4    & 6.3                 & 5.1   & 6.8   & 7.5                     & 15.2                   &  3.5                                \\
  i-\AlTwoOThree      & 3.2                 & -0.9    & -1    & 1.6                & 5.1   & 6.8   & 10.7                    & 59.6                   &  56.7                               \\
  i-\MgO              & 1.1                & -0.7   & -0.8    & 0.6                & 5.1   & 6.8   & 92.4                    & 156.0                  &  377.6                               \\\hline
  \end{tabular}
\caption{Numerical data for the electric double layer formed, respectively, at an intrinsic \SiOTwo, 
\AlTwoOThree, and \MgO\, surface in contact with an hydrogen plasma. The interface is in all three
cases perfectly absorbing as well as collisionless and the plasma is characterized by the parameters 
given in Table~\ref{MaterialParameters}. }
\label{TableForData}
\end{table*}

\end{document}